\documentclass[aps,prd,twocolumn,groupedaddress,amsmath,amssymb]{revtex4-1}
\usepackage{graphicx}  
\usepackage{dcolumn}   
\usepackage{bm}        
\usepackage{verbatim}   
\usepackage{subfigure}
\usepackage[utf8]{inputenc}
\usepackage{color}
\usepackage{braket}

\begin{document}

\title{Production and polarization of prompt $\varUpsilon$($n$S) in the improved color evaporation model using the $k_T$-factorization approach}
\author{Vincent Cheung}
\affiliation{
   Department of Physics,
   University of California, Davis,
   Davis, California 95616, USA
   }
\author{Ramona Vogt}
\affiliation{
   Nuclear and Chemical Sciences Division,
   Lawrence Livermore National Laboratory,
   Livermore, California 94551, USA
   }
\affiliation{
   Department of Physics,
   University of California, Davis,
   Davis, California 95616, USA
   }
\date{\today}
\begin{abstract}
We calculate the polarization of prompt $\varUpsilon$($n$S) production in the improved color evaporation model at leading order employing the $k_T$-factorization approach. We present the polarization parameter $\lambda_\vartheta$ of prompt $\varUpsilon$($n$S) as a function of transverse momentum in $p+p$ and $p+\bar{p}$ collisions to compare with data in the helicity, Collins-Soper and Gottfried-Jackson frames. We also present calculations of the bottomonium production cross sections as a function of transverse momentum and rapidity. This is the first $p_T$-dependent calculation of bottomonium production and polarization in the improved color evaporation model. We find agreement with both bottomonium cross sections and polarization measurements.
\end{abstract}

\pacs{
14.40.Pq
}
\keywords{
Heavy Quarkonia}

\maketitle


\section{Introduction}
This paper is a continuation of our previous work \cite{Cheung:2018tvq} on quarkonium production and polarization in the improved color evaporation model using the $k_T$-factorization approach.

We first developed our LO calculation of quarkonium polarization in the ICEM \cite{Ma:2016exq} in Refs.~\cite{Cheung:2017osx,Cheung:2017loo} employing collinear factorization. However, in this framework, we were unable to address the polarization as a function of $p_T$ to compare with collider data. Therefore, we performed  the first $p_T$-dependent polarization calculation in the ICEM \cite{Cheung:2018tvq} for prompt $J/\psi$ production and polarization by employing the $k_T$-factorization approach. This paper is a continuation of that work where we now extend our $p_T$-dependent leading order (LO) ICEM calculation of quarkonium production and polarization in the $k_T$-factorization approach to prompt $\varUpsilon$($n$S). We use the same scattering amplitudes as in Ref.~\cite{Cheung:2018tvq}. This work also provides the first $p_T$-dependent ICEM $\varUpsilon$($n$S) polarization result. We will begin to address the $p_T$ dependence at NLO in a later publication.

We note that within the framework of Nonrelativistic QCD (NRQCD) \cite{Caswell:1985ui}, the quarkonium polarization problem is less prominent in bottomonium than in charmonium. Fitting the long distance matrix elements to measurements of $\varUpsilon$ yields and polarization for $p_T > 8$~GeV, NRQCD is able to provide a better description of bottomonium yields and polarization than for charmonium \cite{Gong:2012ug,Gong:2013qka}.  The heavier bottom quark mass allows better convergence of the double expansion in $\alpha_s$ and $v$. However, at low $p_T$, NRQCD calculations still overestimate the $\varUpsilon$(1S) experimental yields by a factor of 2 to 3 \cite{Gong:2013qka}.

\section{Production of polarized bottomonium in the $k_T$-factorization approach}

In this paper, we present both the yields and polarizations of bottomonium as a function of $p_T$ by formulating the ICEM in the $k_T$-factorization approach. We take the same effective Feynman rules for scattering processes involving incoming off-shell gluons \cite{Collins:1991ty} as in the NRQCD calculation of Ref.~\cite{Kniehl:2006sk}. Effectively, the momentum of the incoming Reggeon, $k^\mu$, with transverse momentum $k_T$ can be written in terms of the proton momentum $p^\mu$ and the fraction of longitudinal momentum $x$ carried by the gluon as 
\begin{eqnarray}
k^{\mu} = x p^\mu + k_{T}^\mu \;. \label{Reggeon_momentum}
\end{eqnarray}
The polarization 4-vector is
\begin{eqnarray}
\epsilon^\mu(k_{T}) = \frac{k_T^\mu}{k_T} \label{Reggeon_polarization} \;,
\end{eqnarray}
where $k_T^\mu = (0,\vec{k}_T,0)$.

\begin{table}
\caption{\label{states}The mass, $M_\mathcal{Q}$, and the squared feed-down transition Clebsch-Gordan coefficients, $S_\mathcal{Q}^{J_z}$, for all bottomonium states contributing to prompt $\varUpsilon$($n$S) production.}
\begin{ruledtabular}
\begin{tabular}{cccc}
$\mathcal{Q}$ & $M_\mathcal{Q}$ (GeV) & $S_\mathcal{Q}^{J_z=0}$ & $S_\mathcal{Q}^{J_z=\pm1}$\\
\hline
$\varUpsilon$(1S) & 9.46 & 1 & 0 \\
$\varUpsilon$(2S) & 10.02 & 1 & 0 \\
$\varUpsilon$(3S) & 10.36 & 1 & 0 \\
$\chi_{b1}$(1P) & 9.89 & 0 & 1/2 \\
$\chi_{b2}$(1P) & 9.91 & 2/3 & 1/2 \\
$\chi_{b1}$(2P) & 10.26 & 0 & 1/2 \\
$\chi_{b2}$(2P) & 10.27 & 2/3 & 1/2 \\
$\chi_{b1}$(3P) & 10.51 & 0 & 1/2 \\
$\chi_{b2}$(3P) & 10.51 & 2/3 & 1/2 \\
\end{tabular}
\end{ruledtabular}
\end{table}

\begin{table*}
\centering
\caption{\label{feeddowntable} The feed-down ratios, $c_{\mathcal{Q}}$, for prompt $\varUpsilon$(1S), $\varUpsilon$(2S) and $\varUpsilon$(3S) production from direct $\varUpsilon$(1S), $\varUpsilon$(2S), $\varUpsilon$(3S), $\chi_{b}(1P)$, $\chi_{b}(2P)$ and $\chi_{b}(3P)$ in the low $p_T$ and high $p_T$ regions \cite{Andronic:2015wma}. We assume the feed-down contributions from $\chi_{b1}$($n$P) and $\chi_{b2}$($n$P) are the same as also done in Ref.~\cite{Cheung:2017osx}.} \label{feeddown}
\begin{ruledtabular}
\begin{tabular}{c||ccc|ccc}
& & low $p_T$ $c_{\mathcal{Q}}$ ($p_T \lesssim 20$~GeV) & & & high $p_T$ $c_{\mathcal{Q}}$ ($p_T\gtrsim20$~GeV) \\
\hline
$\mathcal{Q}$ (direct~\textbackslash~prompt) & $\varUpsilon$(1S) & $\varUpsilon$(2S) & $\varUpsilon$(3S) & $\varUpsilon$(1S) & $\varUpsilon$(2S) & $\varUpsilon$(3S) \\
\hline
$\varUpsilon$(1S) & 0.71 & - & - & 0.45 & - & - \\
$\varUpsilon$(2S) & 0.07 & 0.73 & - & 0.14 & 0.60 & - \\
$\varUpsilon$(3S) & 0.01 & 0.04 & 0.70 & 0.03 & 0.05 & 0.50 \\
$\chi_{b1}$(1P) & 0.075 & - & - & 0.145 & - & - \\
$\chi_{b2}$(1P) & 0.075 & - & - & 0.145 & - & - \\
$\chi_{b1}$(2P) & 0.02 & 0.10 & - & 0.03 & 0.15 & - \\
$\chi_{b2}$(2P) & 0.02 & 0.10 & - & 0.03 & 0.15 & - \\
$\chi_{b1}$(3P) & 0.01 & 0.015 & 0.15 & 0.015 & 0.025 & 0.25 \\
$\chi_{b2}$(3P) & 0.01 & 0.015 & 0.15 & 0.015 & 0.025 & 0.25 \\
\end{tabular}
\end{ruledtabular}
\end{table*}

In the traditional CEM, all bottomonium states are treated the same as $b\bar{b}$ below the $B\bar{B}$ threshold. The invariant mass of the heavy $b\bar{b}$ pair is restricted to be less than twice the mass of the lowest mass $B$ meson. The distributions for all bottomonium family members are assumed to be identical. In the ICEM, the invariant mass of the intermediate $b\bar{b}$ pair is constrained to be larger than the mass of produced bottomonium state, $M_{\mathcal{Q}}$, instead of twice the bottom quark mass, $2m_b$, the lower limit in the traditional CEM \cite{Barger:1979js,Cheung:2017loo}. Because the bottomonium momentum and integration range now depend on the mass of the state, the kinematic distributions of the bottomonium states are no longer identical in the ICEM. Using the $k_T$-factorization approach, in a $p+p$ collision the ICEM production cross section for a directly-produced bottomonium state $\mathcal{Q}$ is
\begin{widetext}
\begin{eqnarray}
\sigma &=& F_\mathcal{Q} { \int_{ M_\mathcal{Q}^2}^{4m_B^2} d\hat{s} } \int \frac{dx_1}{x_1} \int \frac{d\phi_1}{2\pi} \int {dk_{1T}}^2 \Phi_1(x_1,k_{1T},\mu_{F1}^2) \int \frac{dx_2}{x_2} \int \frac{d\phi_2}{2\pi} \int{dk_{2T}}^2 \Phi_2(x_2,k_{2T},\mu_{F2}^2) \hat{\sigma}(R+R\rightarrow Q\bar{Q}) \nonumber \\
&\times& \delta(\hat{s} - x_1x_2 s +|\vec{k}_{1T}+\vec{k}_{2T}|^2) \;,
\label{cem_sigma}
\end{eqnarray}
\end{widetext}
where the square of the heavy quark pair invariant mass is $\hat{s}$ while the square of the center-of-mass energy in the $p+p$ collision is $s$. Here $\Phi(x,k_{T},\mu_F^2)$ is the unintegrated parton distribution function (uPDF) for a Reggeized gluon with momentum fraction $x$ and transverse momentum $k_T$ interacting with factorization scale $\mu_F$. The angles $\phi_{1,2}$ in Eq.~({\ref{cem_sigma}) are between the $k_{T1,2}$ of the partons and the $p_T$ of the final state bottomonium $\mathcal{Q}$. The parton-level cross section is $\sigma(R+R\rightarrow b\bar{b})$. Finally, $F_{\mathcal{Q}}$ is a universal factor for the directly-produced bottomonium state $\mathcal{Q}$, and is independent of the projectile, target, and energy. In this approach, the cross section is
\begin{widetext}
\begin{eqnarray}
\frac{d^{4}\sigma}{dp_{T} dy d\hat{s} d\phi} &=& \sigma \delta(\hat{s}-x_1 x_2 s + p_T^2) \delta \Big(y-\frac{1}{2}\log\frac{x_1}{x_2} \Big) \delta \Big(p_T^2-|\vec{k}_{1T}^2+\vec{k}_{2T}^2| \Big) \delta(\phi-(\phi_1-\phi_2)) \nonumber \\
&=& F_\mathcal{Q} \int \frac{2}{\pi} k_{2T} dk_{2T} \sum_{k_{1T}} \Bigg[\frac{\Phi_{1}(k_{1T},x_{10},\mu_{F1}^2)}{x_{10}} \frac{\Phi_{2}(k_{2T},x_{20},\mu_{F2}^2)}{x_{20}} k_{1T} p_T \frac{\hat{\sigma}(R+R\rightarrow Q\bar{Q})}{s \sqrt{k_{2T}^2(\cos^2\phi-1)+p_T^2}} \Bigg] \;
\label{int_rapidity}
\end{eqnarray}
\end{widetext}

where the sum $k_{1T}$ is over the roots of $k_{1T}^2+k_{2T}^2+2k_{1T}k_{2T}\cos\phi=p_T^2$, and $k_{1T,1}$, $k_{1T,2}$ are
\begin{eqnarray}
k_{1T,1} &=& -k_{2T}\cos \phi +\sqrt{k_{2T}^2(\cos^2\phi-1)+p_T^2} \label{k11} \\
k_{1T,2} &=& -k_{2T}\cos \phi -\sqrt{k_{2T}^2(\cos^2\phi-1)+p_T^2} \label{k12} \;.
\end{eqnarray}
The momentum fractions $x_{10}$ and $x_{20}$ are 
\begin{eqnarray}
x_{10} &=& \sqrt{\frac{\hat{s}+p_T^2}{s}}e^{+y} \;, \\
x_{20} &=& \sqrt{\frac{\hat{s}+p_T^2}{s}}e^{-y} \;.
\end{eqnarray}
Here, $\phi$ is the relative azimuthal angle between two incident Reggeons ($\phi=\phi_1-\phi_2$) and $p_T$ is the transverse momentum of the produced $b\bar{b}$.

\begin{figure}[hb]
\centering
\includegraphics[width=\columnwidth]{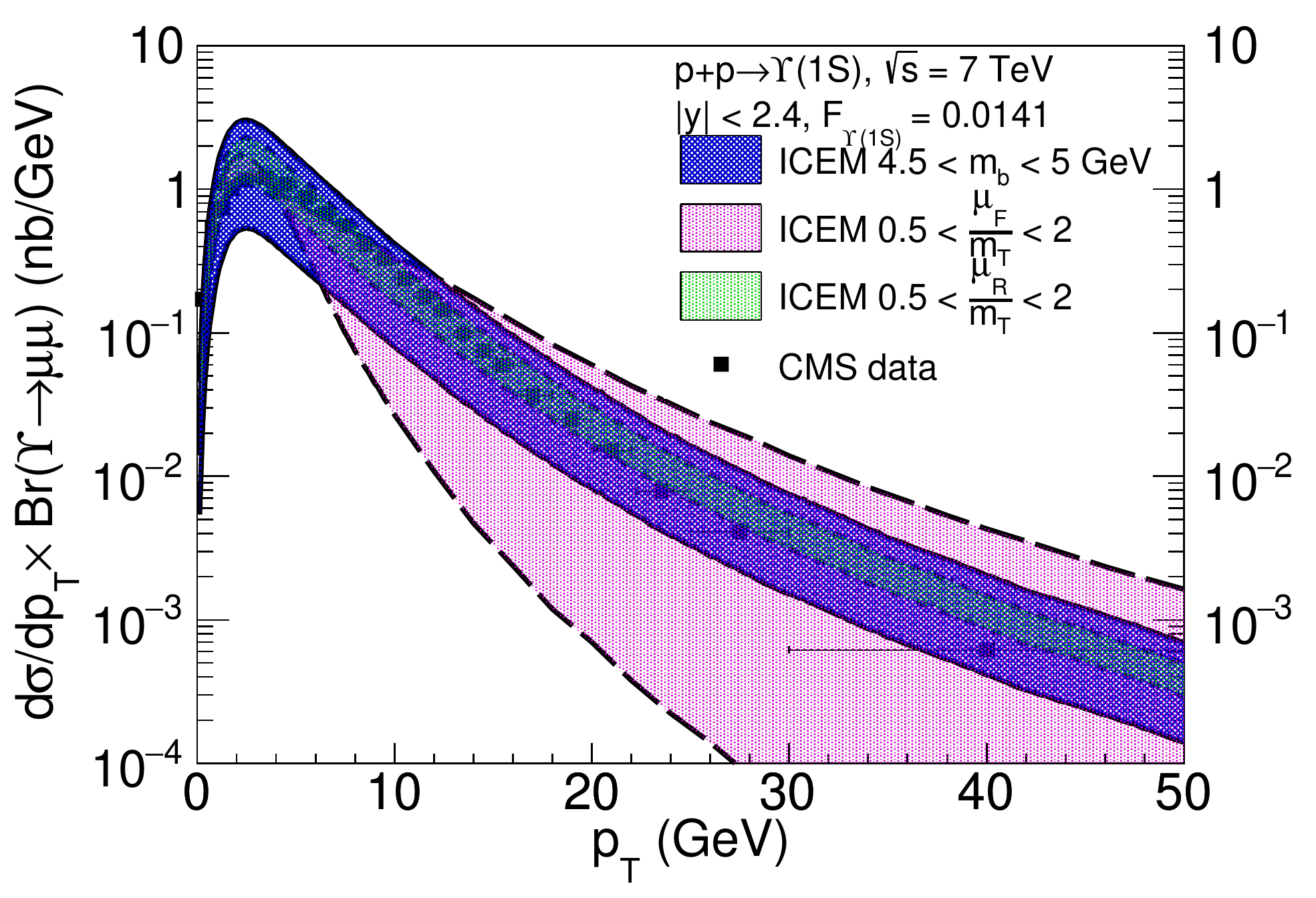}
\caption{(Color online) The $p_T$ dependence of prompt $\varUpsilon$(1S) production at $\sqrt{s} = 7$~TeV in the ICEM obtained by varying the bottom quark mass (blue), the factorization scale in the range $0.5<\mu_F/m_T<2$ (magenta), and the renormalization scale in the range $0.5<\mu_R/m_T<2$ (green) is compared with the CMS midrapidity data \cite{Chatrchyan:2013yna}.} \label{CMS_1S_pt_all_variations}
\end{figure}

Thus the transverse momentum distribution $d\sigma/dp_T$ in the ICEM is
\begin{eqnarray}
\label{cem_pt}
\frac{d\sigma}{dp_T} &=& \int dy d\hat{s} d\phi \frac{d^{4}\sigma}{dp_{T} dy d\hat{s} d\phi} \;. \label{pt_rap_cut}
\end{eqnarray}
We integrate over rapidity to compare to collider data with defined rapidity cuts. Similarly, the rapidity distribution $d\sigma/dy$ in the ICEM is
\begin{eqnarray}
\label{cem_y}
\frac{d\sigma}{dy} &=& \int dp_T d\hat{s} d\phi \frac{d^{4}\sigma}{dp_{T} dy d\hat{s} d\phi} \;. \label{y_pt_cut}
\end{eqnarray}

As our central result, we take the renormalization and factorization scales to be $\mu_F=\mu_R=m_T$, where $m_T$ is the transverse mass of the $b\bar{b}$. We will study the effect of varying these scales on the $p_T$ distributions and the polarization.

\section{Polarization of prompt $\varUpsilon$($n$S)}

We employ the scattering amplitudes calculated in Ref.~\cite{Cheung:2018tvq} to compute the $b\bar{b}$ partonic production cross section $\hat{\sigma}^{J,J_z}$ according to the $J^{P}$ of each directly produced bottomonium state below the $B\bar{B}$ threshold. We then convolute the polarized partonic cross sections with the uPDFs to obtain the hadron-level cross section, $\sigma$, as a function of $p_T$ using Eq.~(\ref{pt_rap_cut}). The bottomonium masses which appear as the lower limit of the $b\bar{b}$ invariant mass in the calculations of $\hat{\sigma}^{J,J_z}$ are listed in Table~\ref{states}. We employ the ccfm-JH-2013-set1 \cite{Hautmann:2013tba} uPDFs in this calculation.

We assume that the angular momentum of each directly-produced bottomonium state is unchanged by the transition from the parton level to the hadron level, consistent with the CEM expectation that the linear momentum is unchanged by hadronization.

\begin{figure*}
\centering
\begin{minipage}[ht]{0.97\columnwidth}
\centering
\includegraphics[width=\columnwidth]{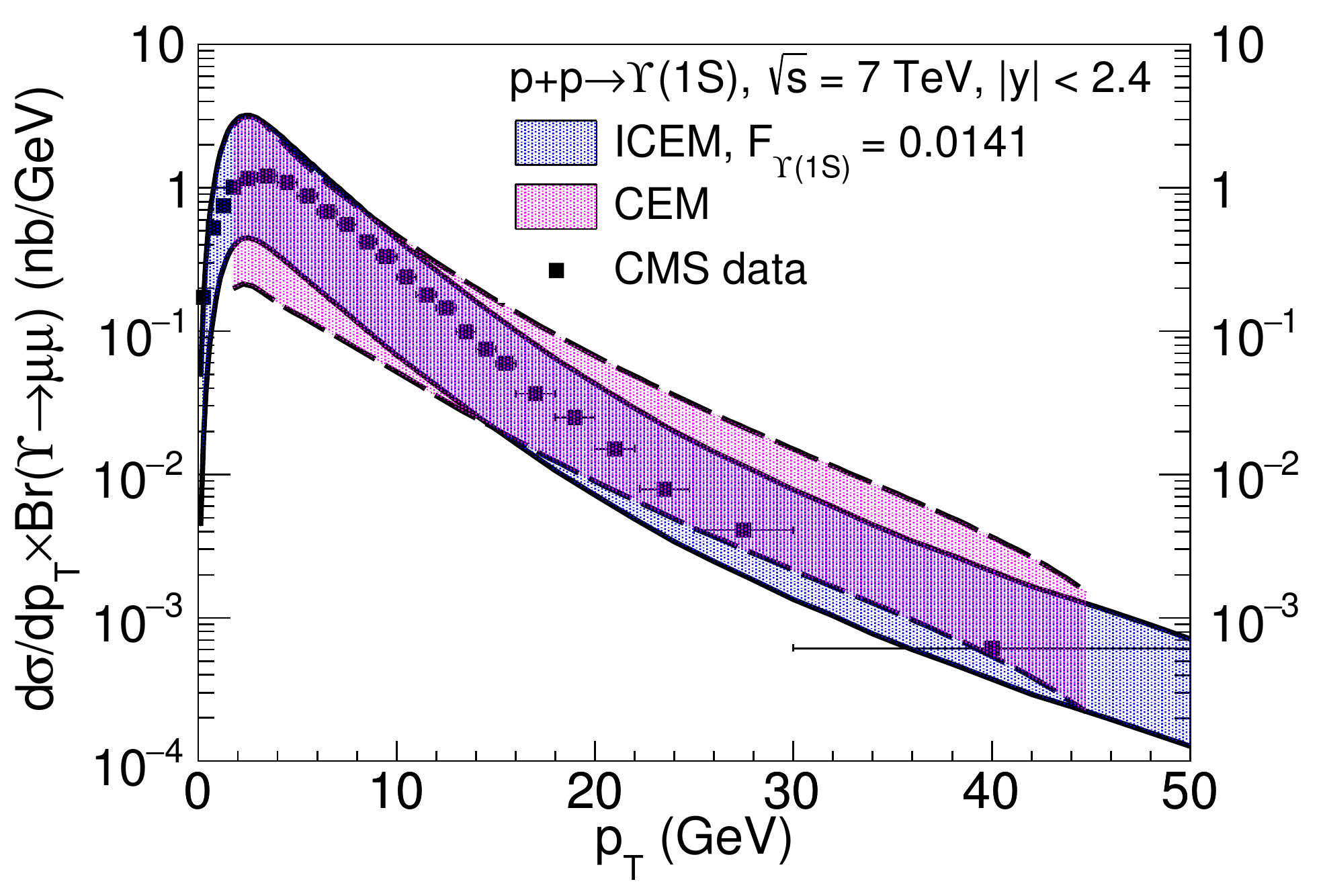}
\caption{(Color online) The $p_T$ dependence of prompt $\varUpsilon$(1S) production at $\sqrt{s} = 7$~TeV in the ICEM with combined mass and renormalization scale uncertainties (blue) and that in the CEM using collinear factorization approach (magenta). The CMS midrapidity data \cite{Chatrchyan:2013yna} from Fig.~\ref{CMS_1S_pt_all_variations} are also shown.} \label{CMS_1S_pt}
\end{minipage}%
\hspace{1cm}%
\begin{minipage}[ht]{0.97\columnwidth}
\centering
\includegraphics[width=\columnwidth]{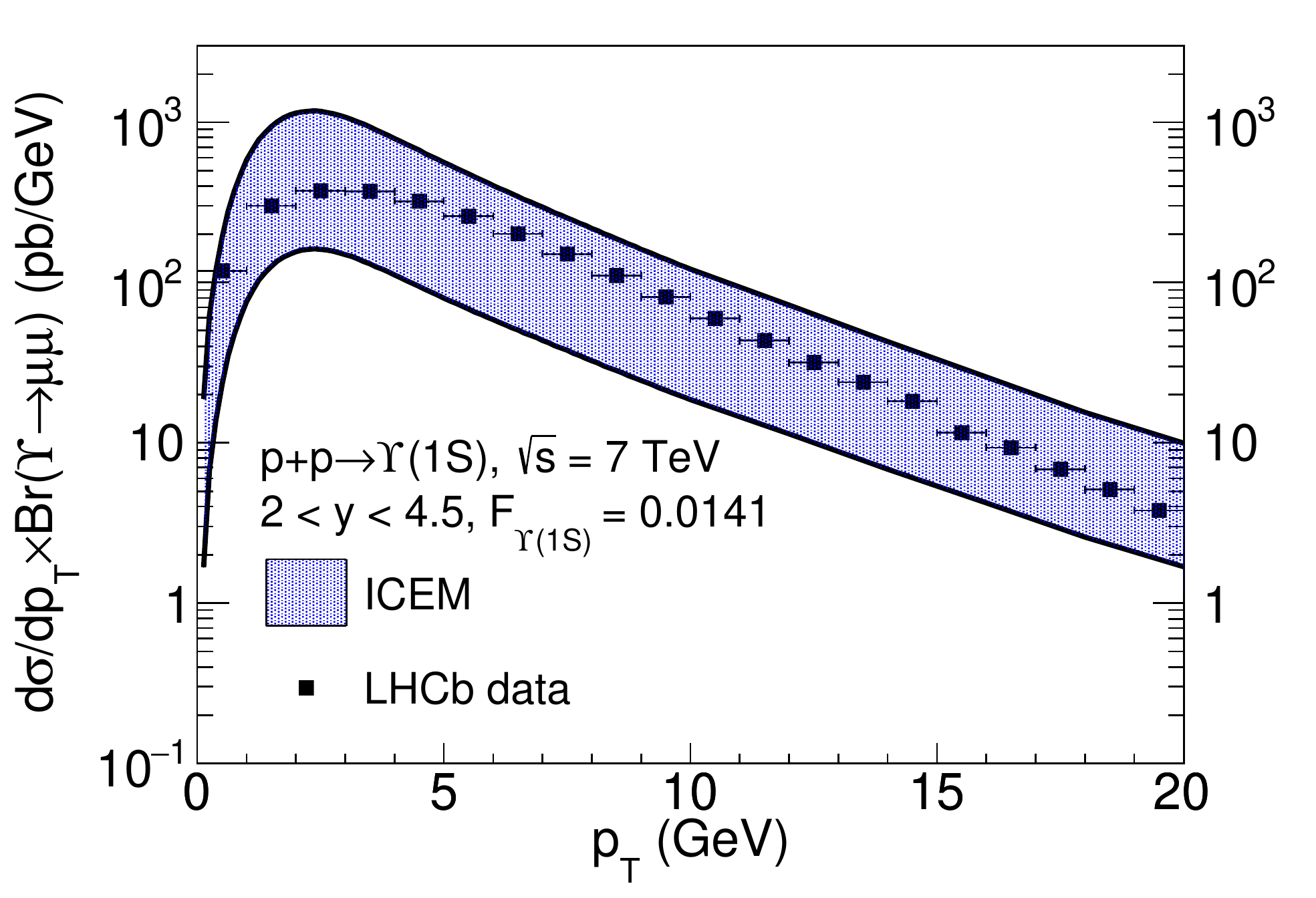}
\caption{The $p_T$ dependence of prompt $\varUpsilon$(1S) production at $\sqrt{s} = 7$~TeV and $2<y<4.5$ in the ICEM with combined mass and renormalization scale uncertainties is compared with the LHCb data \cite{Aaij:2015awa}.} \label{LHCb_1S_pt}
\end{minipage}
\end{figure*}

\begin{figure}[hb]
\centering
\includegraphics[width=\columnwidth]{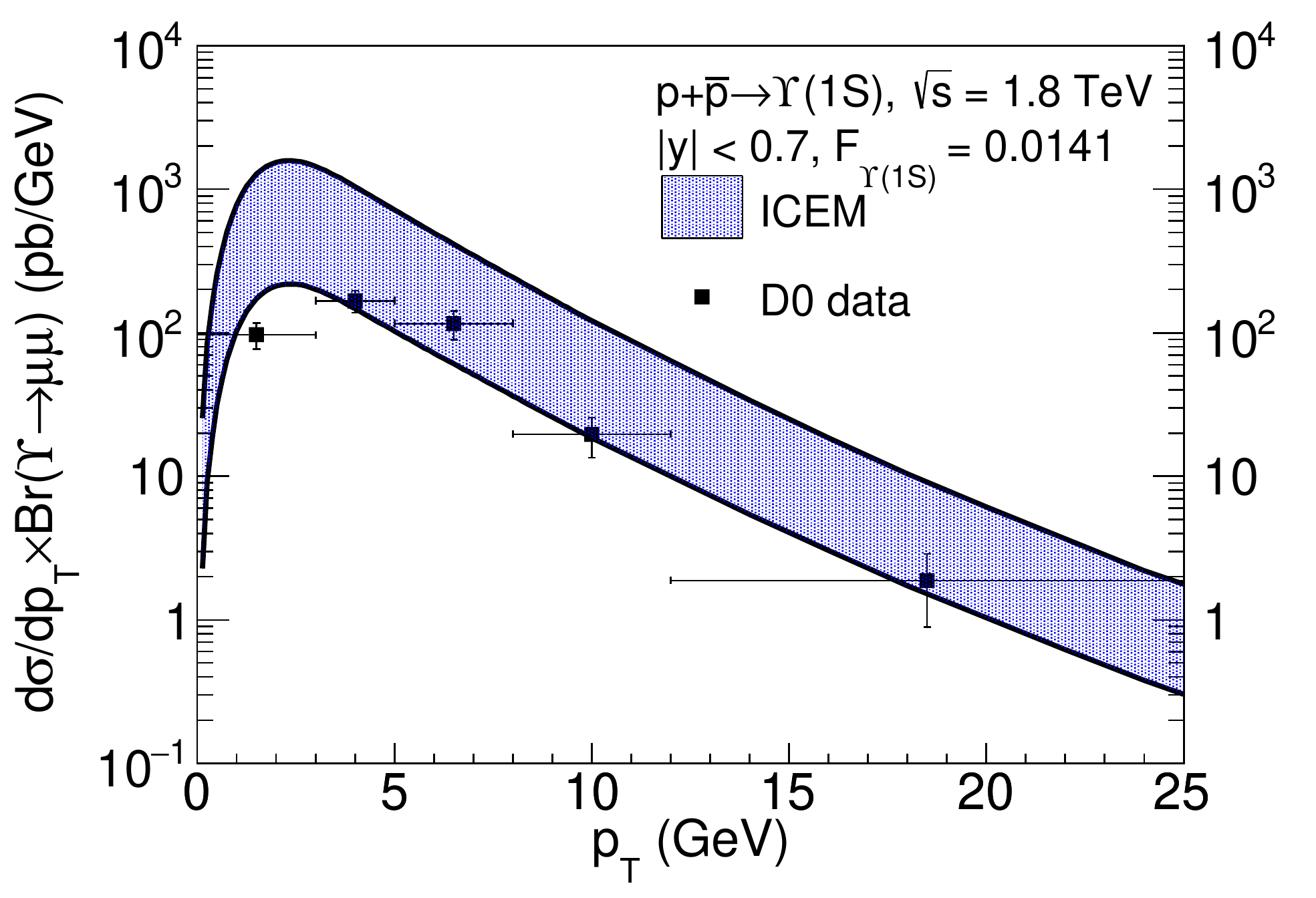}
\caption{The $p_T$ dependence of prompt $\varUpsilon$(1S) production at $\sqrt{s} = 7$~TeV and $|y|<0.7$ in the ICEM with combined mass and renormalization scale uncertainties is compared with the D0 data \cite{Abachi:1995tj}.} \label{D0_1S_pt}
\end{figure}

We calculate the ratio of the individual $J_z=0,\pm1$ to the unpolarized partonic cross sections ratios for each directly-produced bottomonium state $\mathcal{Q}$ that has a contribution to prompt $\varUpsilon$($n$S) production: $\varUpsilon$(1S), $\varUpsilon$(2S), $\varUpsilon$(3S), $\chi_{b1}$(1P), $\chi_{b2}$(1P), $\chi_{b1}$(2P), $\chi_{b2}$(3P), $\chi_{b1}$(3P) and $\chi_{b2}$(3P). These ratios, $R_{\mathcal{Q}}^{J_z}$, are then independent of $F_{\mathcal{Q}}$. We assume the feed-down production of $\varUpsilon$($n$S) from the higher mass bound states follows the angular momentum algebra. Their contributions of these higher states to $R_{\varUpsilon{\rm(}n{\rm S)}}^{J_z=0}$ for prompt $\varUpsilon$($n$S) are added after weighting by the feed-down contribution ratios $c_{\mathcal{Q}}$ \cite{Andronic:2015wma}:
			
\begin{eqnarray}
\label{mix_upsilon}
R_{\varUpsilon}^{J_z=0} &=& \sum_{\mathcal{Q},J_z} c_{\mathcal{Q}} S_{\mathcal{Q}}^{J_z} R_{\mathcal{Q}}^{J_z} \;.
\end{eqnarray}
Here $S_{\mathcal{Q}}^{J_z}$ is the transition probability from a given state $\mathcal{Q}$ produced in a $J_z$ state to a $\varUpsilon$($n$S) with $J_z=0$ in a single decay. We assume two pions are emitted for S state feed down, $\varUpsilon{\rm (2S)}\rightarrow \varUpsilon{\rm(1S)}\pi \pi$, and a photon is emitted for a P state feed down, $\chi_{b}({\rm 1P})\rightarrow \varUpsilon({\rm 1S}) \gamma$. $S_{\mathcal{Q}}^{J_z}$ is then 1 (if $J_z=0$) or 0 (if $J_z=1$) for $\mathcal{Q}=\varUpsilon$(2S) since the transition, $\varUpsilon{\rm (2S)} \rightarrow \varUpsilon{\rm(1S)}\pi \pi$, does not change the angular momentum of the quarkonium state. For directly produced $\varUpsilon$($n$S), $S_{\mathcal{Q}}^{J_z}$ is 1 for $J_z=0$ and 0 for $J_z=1$. The $S_{\mathcal{Q}}^{J_z}$ for the $\chi$ states are the squares of the Clebsch-Gordan coefficients for the feed-down production via $\chi_b \rightarrow \varUpsilon(n{\rm S})\gamma$. The bottomonium feed-down ratios are $p_T$-dependent \cite{Andronic:2015wma}: the fraction of direct production is larger at low $p_T$ than at high $p_T$. We consider two sets of feed-down ratios from Ref.~\cite{Andronic:2015wma}. These ratios are derived from LHC measurements \cite{Aad:2012dlq,Chatrchyan:2013yna,Abelev:2014qha,Aaij:2014caa,Aad:2011xv,Khachatryan:2010zg,Aaij:2013yaa,Aaij:2012se,LHCb:2012aa} assuming they vary with $p_T$ but not rapidity \cite{Andronic:2015wma}. The ``low $p_T$'' ratios are used to compare with LHCb data ($0<p_T<20$~GeV) where the ``high $p_T$'' ratios are employed to compare with CMS data ($10<p_T<50$~GeV). Here, we are assuming the feed-down contribution from $\chi_{b1}$($n$P) and $\chi_{b2}$($n$P) are the same as in our previous approach for the $\chi_c$ states \cite{Cheung:2017osx}. A similar assumption is made for the other P states. The values of $M_\mathcal{Q}$ and $S_\mathcal{Q}^{J_z}$ for all bottomonium states contributing to prompt $\varUpsilon$($n$S) production are collected in Table~\ref{states} and the values of $c_\mathcal{Q}$ in the two $p_T$ regions are presented in Table~\ref{feeddown}.

Finally, the $J_z=0$ to the unpolarized ratio for prompt $\varUpsilon$($n$S) states are converted into the polarization parameter $\lambda_\vartheta$ \cite{Faccioli:2010kd},
\begin{eqnarray}
\label{s_state_lambda}
\lambda_{\vartheta} &=& \frac{1-3R^{J_z=0}}{1+R^{J_z=0}} \;,
\end{eqnarray}
where $-1<\lambda_\vartheta<1$. If $\lambda_\vartheta = -1$, $\varUpsilon$($n$S) production is totally longitudinal, $\lambda_\vartheta= 0$ refers to unpolarized production, while production is totally transverse for $\lambda_\vartheta=+1$.


\section{Results}

Although the matrix elements in this calculation are LO in $\alpha_s$, by convoluting the polarized partonic cross sections with the transverse momentum dependent uPDFs using the $k_T$-factorization approach, we can calculate the yield as well as the polarization parameter $\lambda_\vartheta$ as a function of $p_T$. The full NLO polarization, including $q\bar{q}$ and $(q+\bar{q})g$ contributions, will be discussed in a future publication.

The traditional CEM can describe the unpolarized yields of $\varUpsilon$($n$S) production at NLO assuming collinear factorization \cite{Schuler:1996ku}. In this calculation, we take advantage of the ICEM to calculate the direct production of the individual bottomonium states separately. Since this is the first bottomonium calculation in the ICEM using the $k_T$-factorization approach, it is important to check if our calculated unpolarized yields are also in agreement with the data.

\begin{figure*}
\centering
\begin{minipage}[ht]{0.97\columnwidth}
\centering
\includegraphics[width=\columnwidth]{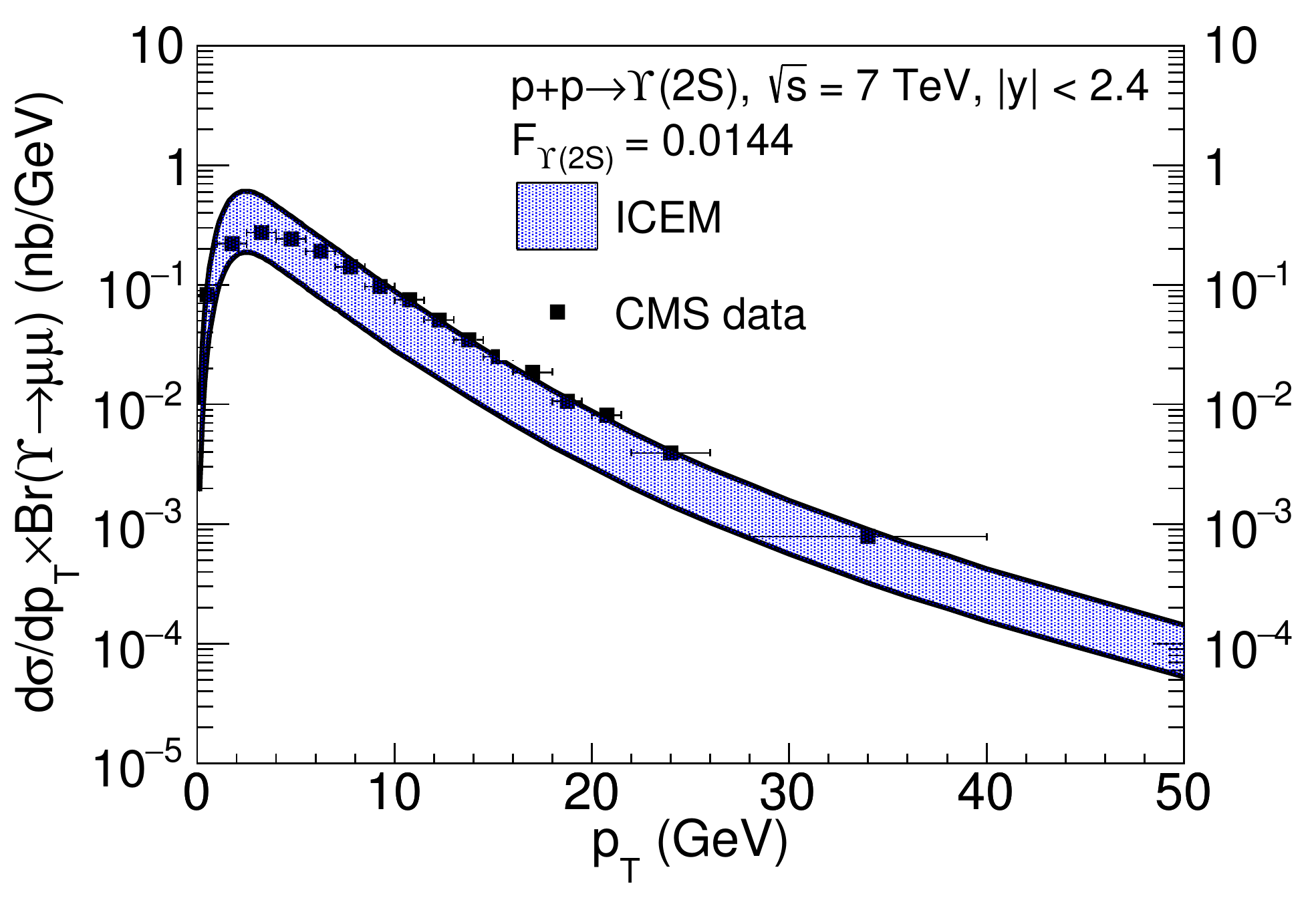}
\caption{The $p_T$ dependence of prompt $\varUpsilon$(2S) production at $\sqrt{s} = 7$~TeV and $2<y<4.5$ in the ICEM with combined mass and renormalization scale uncertainties is compared with the CMS midrapidity data \cite{Chatrchyan:2013yna}.} \label{CMS_2S_pt}
\end{minipage}%
\hspace{1cm}%
\begin{minipage}[ht]{0.97\columnwidth}
\centering
\includegraphics[width=\columnwidth]{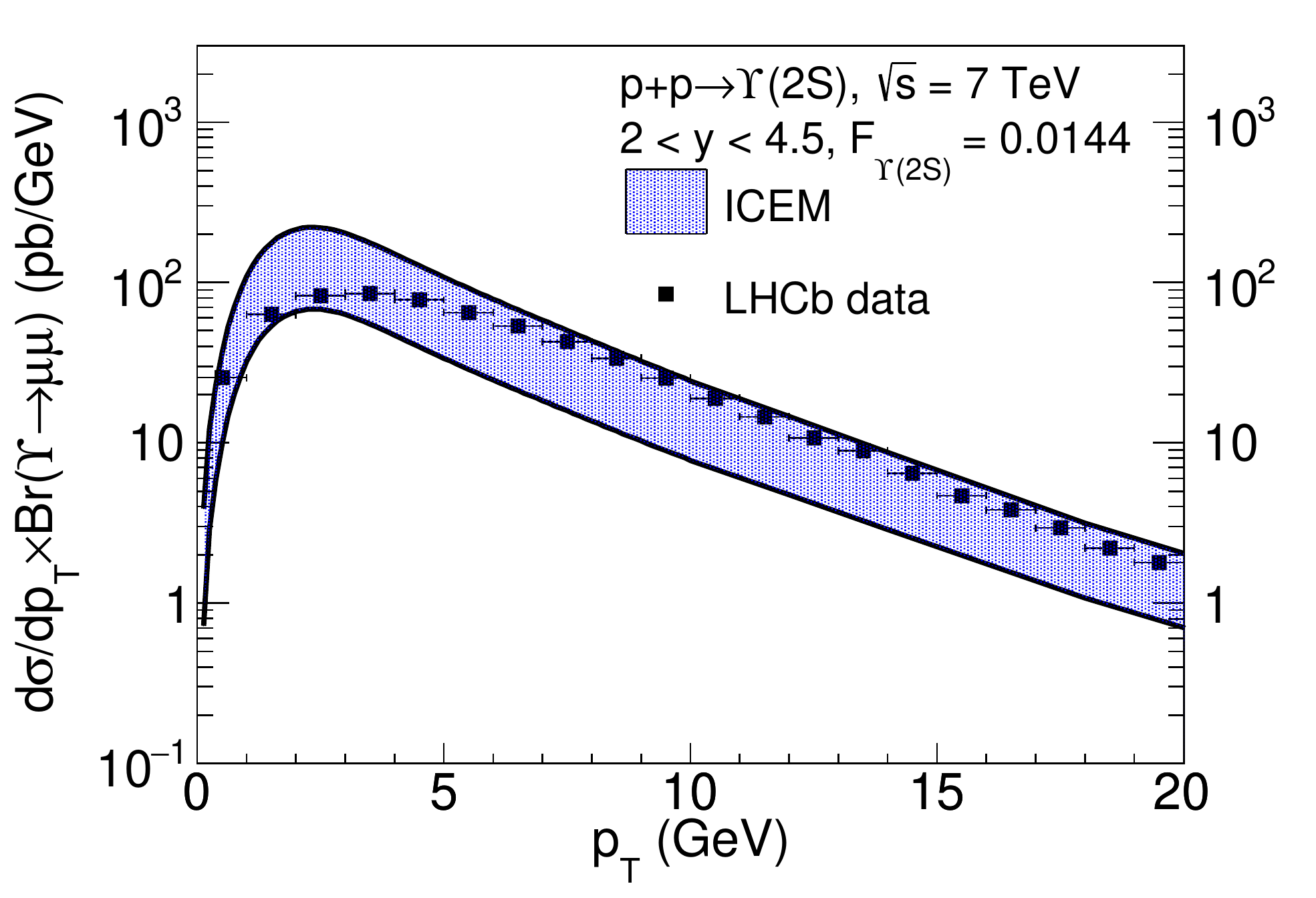}
\caption{The $p_T$ dependence of prompt $\varUpsilon$(2S) production at $\sqrt{s} = 7$~TeV and $2<y<4.5$ in the ICEM with combined mass and renormalization scale uncertainties is compared with the LHCb data \cite{Aaij:2015awa}.} \label{LHCb_2S_pt}
\end{minipage}
\end{figure*}

We first check how our approach describes the transverse momentum and rapidity distribution of the bottomonium states at collider energies. We then discuss the transverse momentum dependence of the polarization parameter $\lambda_\vartheta$ for prompt $\varUpsilon$($n$S) production. We compare our results to the polarization measured in collider experiments in the helicity (HX), Collins-Soper (CS) \cite{Collins:1977iv}, and Gottfried-Jackson (GJ) \cite{Gottfried:1964nx} frames to discuss the frame dependence of $\lambda_\vartheta$. W also discuss the sensitivity of our results to the bottom quark mass, the renormalization scale and the feed-down ratios. In our calculations, we construct the uncertainty bands by varying the bottom quark mass around its base value of 4.75~GeV, in the interval $4.5 < m_b < 5$~GeV, and the renormalization scale around its base value of $m_T$, in the interval $0.5 < \mu_R/m_T < 2$, while keeping the factorization scale fixed at $\mu_F=m_T$. The total uncertainty band is constructed by adding the mass and renormalization scale uncertainties in quadrature. We do not extend our calculation below $p$+$\bar{p}$ at Tevatron energies because at fixed-target energies and even at the RHIC collider the $k_T$-factorization approach with off-shell gluons is inappropriate for bottomonium.
\subsection{Unpolarized bottomonium production}

Here, we present the $p_T$ and rapidity distributions of the $\varUpsilon$($n$S) states as well as the ratio of $\chi_{b1}$(1P) to $\chi_{b2}$(1P) in our approach. In the spirit of the traditional CEM, $F_\mathcal{Q}$ in Eq.~(\ref{cem_sigma}) has to be independent of the projectile, target, and energy for each bottomonium state $\mathcal{Q}$. Even though the focus of this paper is on polarization, independent of $F_\mathcal{Q}$, the unpolarized bottomonium yields in the ICEM using the $k_T$-factorization approach were not calculated before. Therefore, it is important to first confirm that this approach can indeed describe the bottomonium yields as a function of $p_T$ and rapidity before discussing polarization. The direct production cross section is calculated using Eq.~(\ref{pt_rap_cut}) by integrating the pair invariant mass from $M_{\mathcal{Q}}$ to $2m_{B^0}$ ($m_{B^0}=5.28$~GeV). 

We first obtain $F_{\varUpsilon(n{\rm S})}$ by comparing our results with the $\varUpsilon$($n$S) yields measured by the CMS Collaboration at 7~TeV. Using the same $F_{\varUpsilon(n{\rm S})}$, we compare our results with the $\varUpsilon$($n$S) data measured at CDF and LHCb.

\begin{figure*}
\centering
\begin{minipage}[ht]{0.97\columnwidth}
\centering
\includegraphics[width=\columnwidth]{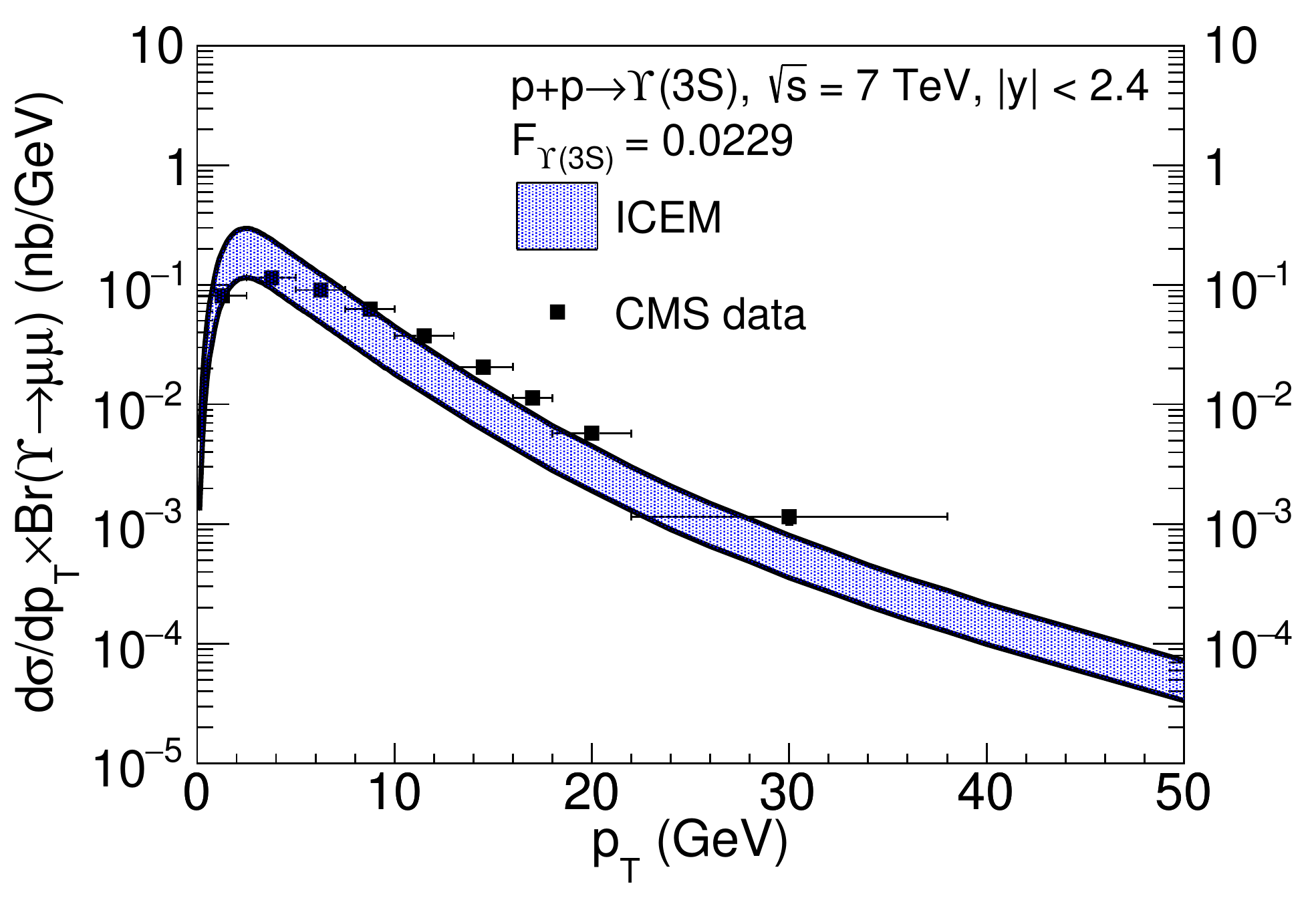}
\caption{The $p_T$ dependence of prompt $\varUpsilon$(3S) production at $\sqrt{s} = 7$~TeV in the ICEM with combined mass and renormalization scale uncertainties is compared with the CMS midrapidity data \cite{Chatrchyan:2013yna}.} \label{CMS_3S_pt}
\end{minipage}%
\hspace{1cm}%
\begin{minipage}[ht]{0.97\columnwidth}
\centering
\includegraphics[width=\columnwidth]{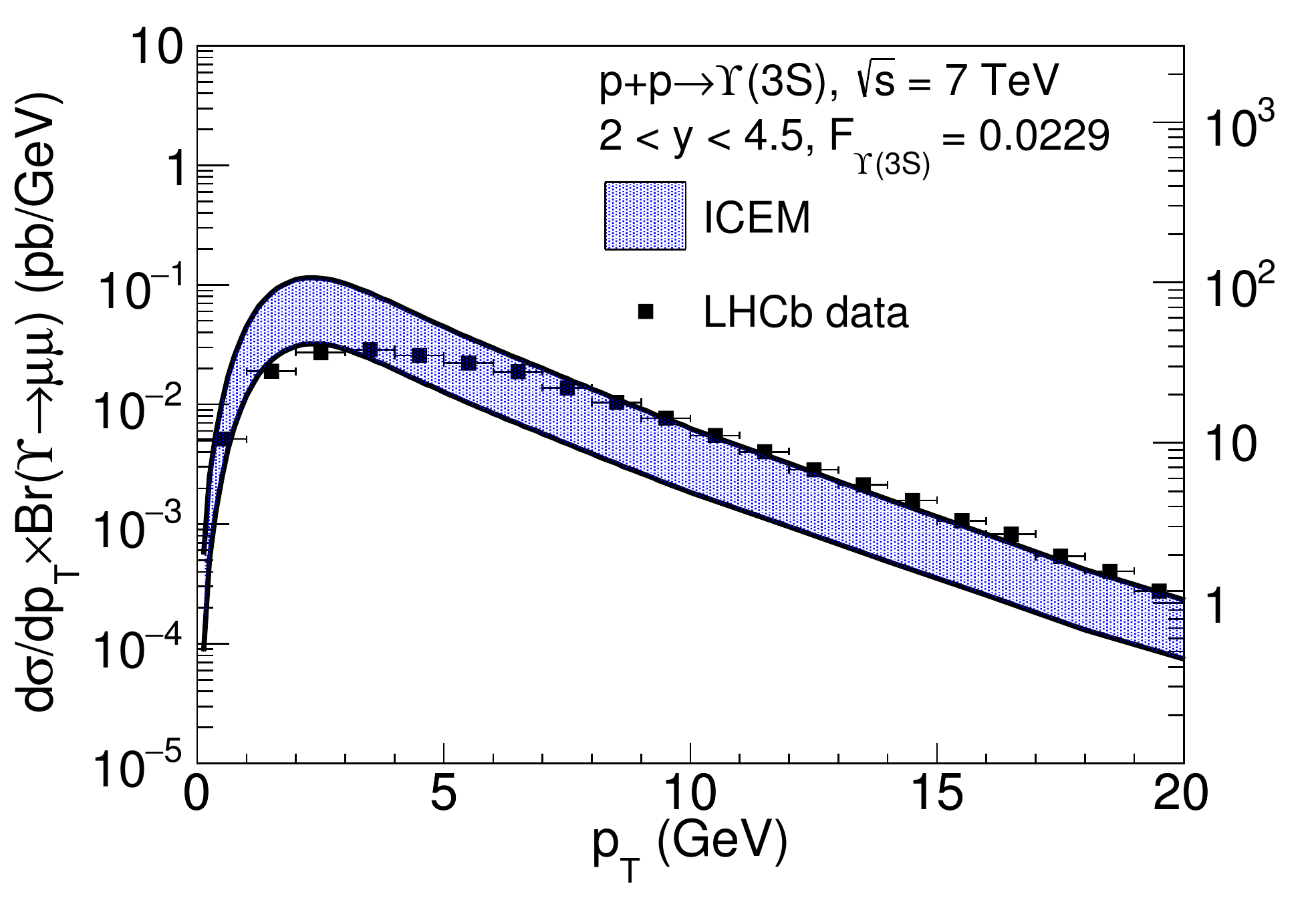}
\caption{The $p_T$ dependence of prompt $\varUpsilon$(2S) production at $\sqrt{s} = 7$~TeV and $2<y<4.5$ in the ICEM with combined mass and renormalization scale uncertainties is compared with the LHCb data \cite{Aaij:2015awa}.} \label{LHCb_3S_pt}
\end{minipage}
\end{figure*}

\subsubsection{$\varUpsilon$(1S) $p_T$ distribution} \label{factorization}
We found in our previous paper \cite{Cheung:2018tvq} that the charmonium $p_T$ distribution has a significant dependence on the factorization scale for $p_T>5$~GeV. In this paper, we also fix the factorization scale at $\mu_F=m_T$ instead of including a factor of two variation. In Fig.~\ref{CMS_1S_pt_all_variations}, we show the $p_T$ distributions of prompt $\varUpsilon$(1S) production at $\sqrt{s}=7$~TeV found by fixing $m_b = 4.75$~GeV and varying the factorization scale over the range $0.5 < \mu_F/m_T < 2$ and the renormalization scale over the range $0.5 < \mu_R/m_T < 2$ separately. We also fix $\mu_F/m_T=\mu_R/m_T=1$ and vary the bottom quark mass over the range $4.5<m_b<5$~GeV. The direct production cross section is calculated using Eq.~(\ref{pt_rap_cut}) by integrating the pair invariant mass from $M_{\varUpsilon{\rm(1S)}}$ to $2m_{B^0}$ ($m_{D^0}=5.28$~GeV) over the rapidity range $|y| < 2.4$. We assume that direct production is a constant fraction, $0.71$ of the prompt production, according to the low $p_T$ feed-down coefficients in Table~\ref{feeddown}, since the yield is dominated by production at low $p_T$. We then compare the prompt $p_T$ distribution in the ICEM with the CMS data \cite{Chatrchyan:2013yna}. Similar to the charmonium $p_T$ distribution, the result has a significant dependence on the factorization scale for $p_T>5$~GeV. This is because the uPDFs have a sharp cutoff for $k_T > \mu_F$ and are thus very sensitive to the chosen factorization scale. The yield varies more as $p_T$ approaches $m_T$ at high $p_T$. At low $p_T$, $m_T \sim M_\mathcal{Q}$ and the cross section is independent of the factorization scale since $k_T \ll \mu_F$. At moderate $p_T$, the variation with $\mu_F$ is similar to or smaller than that due to the bottom quark mass. At $p_T\sim10$~GeV, $m_T \sim p_T$. Thus the lower limit on the factorization scale, $m_T/2$, is on the order of $k_T$ and the yield drops off at this cutoff limit of $\sim5$~GeV, while the upper limit on the factorization scale, $2m_T$, is still greater than $k_T$, enhancing the yield. Since, at LO, only the $b\bar{b}$ pair carries the transverse momentum, the predictive power for the yields is limited by the uPDFs. Therefore, to construct a meaningful uncertainty band, we fix the factorization scale at $\mu_F=m_T$. As we push toward the limit of the $k_T$-factorization approach with uPDFs at high $p_T$ at LO, we can only improve the high $p_T$ limit by a full NLO calculation in the collinear factorization approach where there is no hard limit on $\mu_F$ as in $k_T$-factorization approach.

After fixing the factorization scale, the variation in bottom quark mass then gives the largest uncertainty, followed by the variation in renormalization scale. When $\mu_R$ is reduced, the strong coupling constant is larger, increasing the yield. On the other hand, when $m_b$ is reduced, the yield increases. In the remainder of this section, we present our results by adding the uncertainties due to variations of the bottom mass and renormalization scale in quadrature. 

The prompt $\varUpsilon$(1S) $p_T$ distribution at $\sqrt{s} = 7$~TeV with combined uncertainty is shown in Fig.~\ref{CMS_1S_pt}. The ICEM result has a peak at $p_T$ $\sim$ 2.5~GeV, in agreement with the data. By matching to the total experimental unpolarized yield in $|y|<2.4$, we find that the ICEM can describe the $\varUpsilon$(1S) $p_T$ distribution with $F_{\varUpsilon{\rm(1S)}}=0.0141$. This is the fraction of $b\bar{b}$ pairs produced in the invariant mass range from $M_{\varUpsilon(1S)}$ to $2m_{B^0}$, a difference of $\sim$1~GeV, that result in direct $\varUpsilon$(1S) production, defined in Eq.~(\ref{cem_sigma}). In general, the ICEM $p_T$ distribution agrees with the data for all $p_T$.

In the same figure, we compare the inclusive $\varUpsilon$(1S) $p_T$ distributions with that from the CEM in the collinear factorization approach. The uncertainty band is constructed by combining the uncertainty by varying the bottom mass in the range $4.56<m_b<4.74$~GeV, the factorization scale in the range $0.91<\mu_F/m_T<2.17$, and the renormalization scale in the range $0.9<\mu_R/m_T<1.32$. We find two distributions agree reasonably well with each other and the data.

\begin{figure*}
\centering
\begin{minipage}[ht]{0.97\columnwidth}
\centering
\includegraphics[width=\columnwidth]{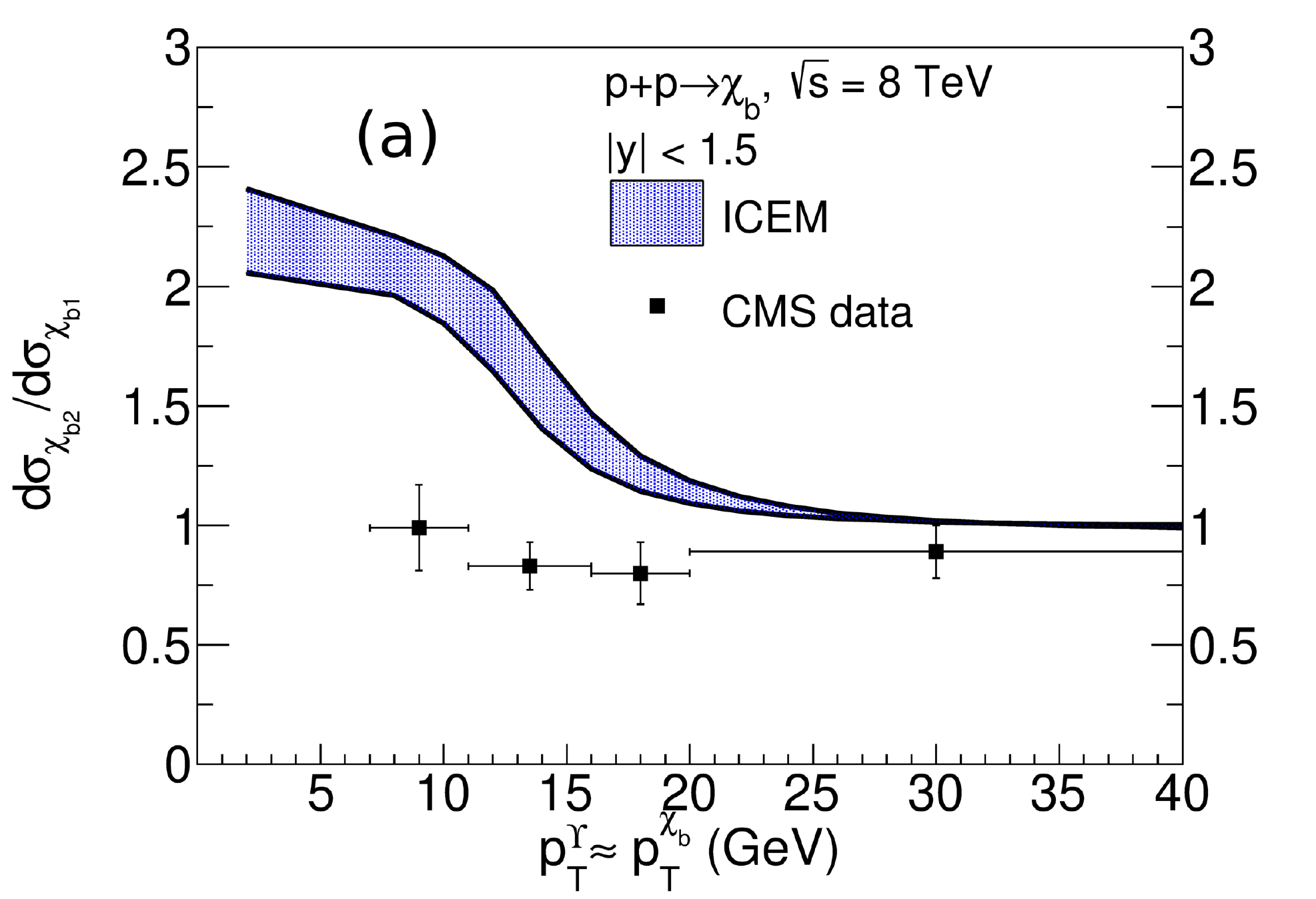}
\end{minipage}
\begin{minipage}[ht]{0.97\columnwidth}
\centering
\includegraphics[width=\columnwidth]{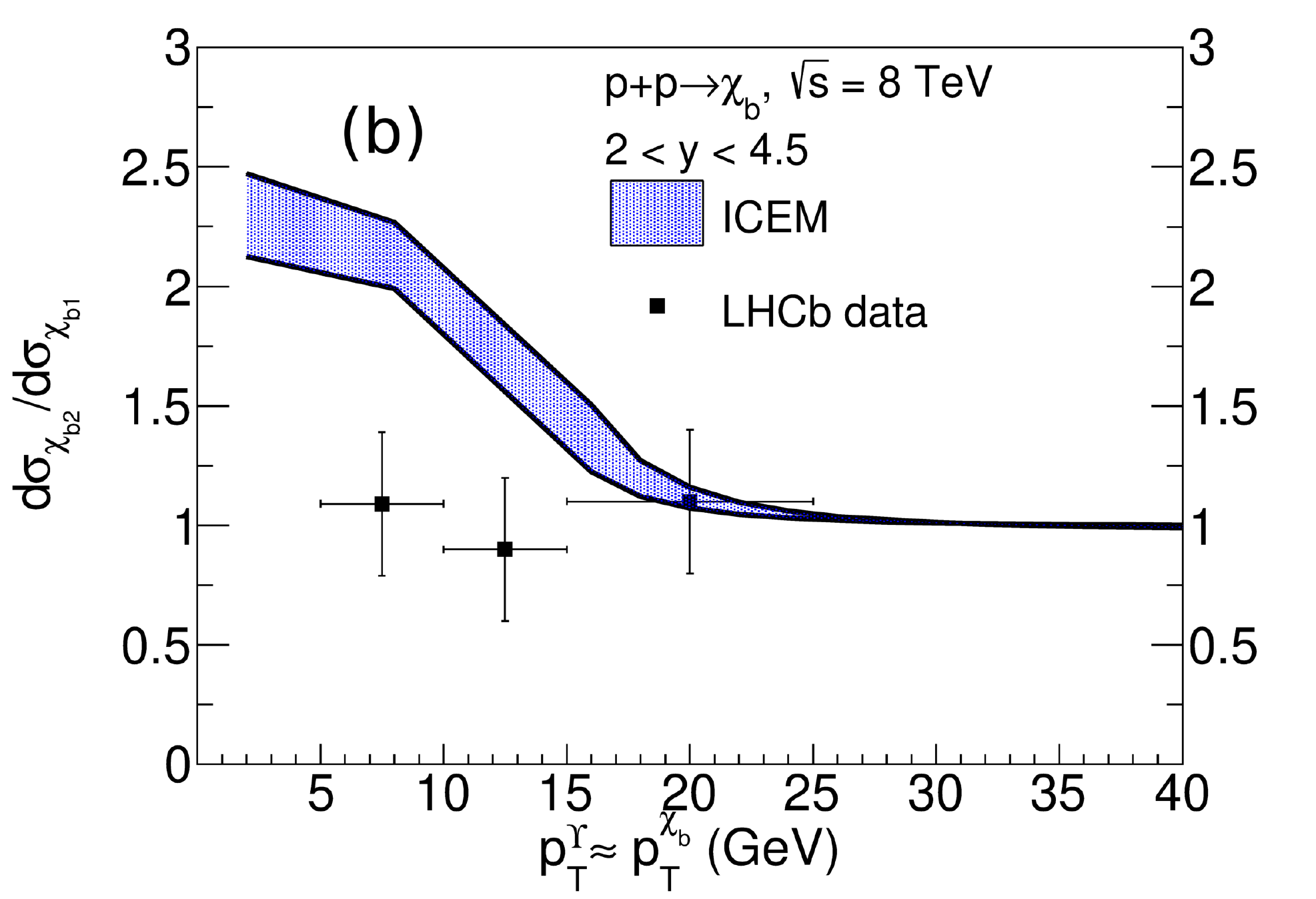}
\end{minipage}
\caption{The ratio of $\chi_{b2}$(1P) to $\chi_{b1}$(1P) in the ICEM with combined mass and renormalization scale uncertainties at $\sqrt{s}= 8$~TeV at central rapidity $|y|<1.5$ (a) and at forward rapidity $2<y<4.5$ (b) assuming $F_{\chi_{b1{\rm(1P)}}}=F_{\chi_{b2{\rm(1P)}}}$. The CMS data \cite{Khachatryan:2014ofa} and the LHCb data \cite{Aaij:2014hla} are also shown in (a) and (b) respectively.} \label{chib_ratios}
\end{figure*}
\begin{figure}[hb]
\centering
\includegraphics[width=\columnwidth]{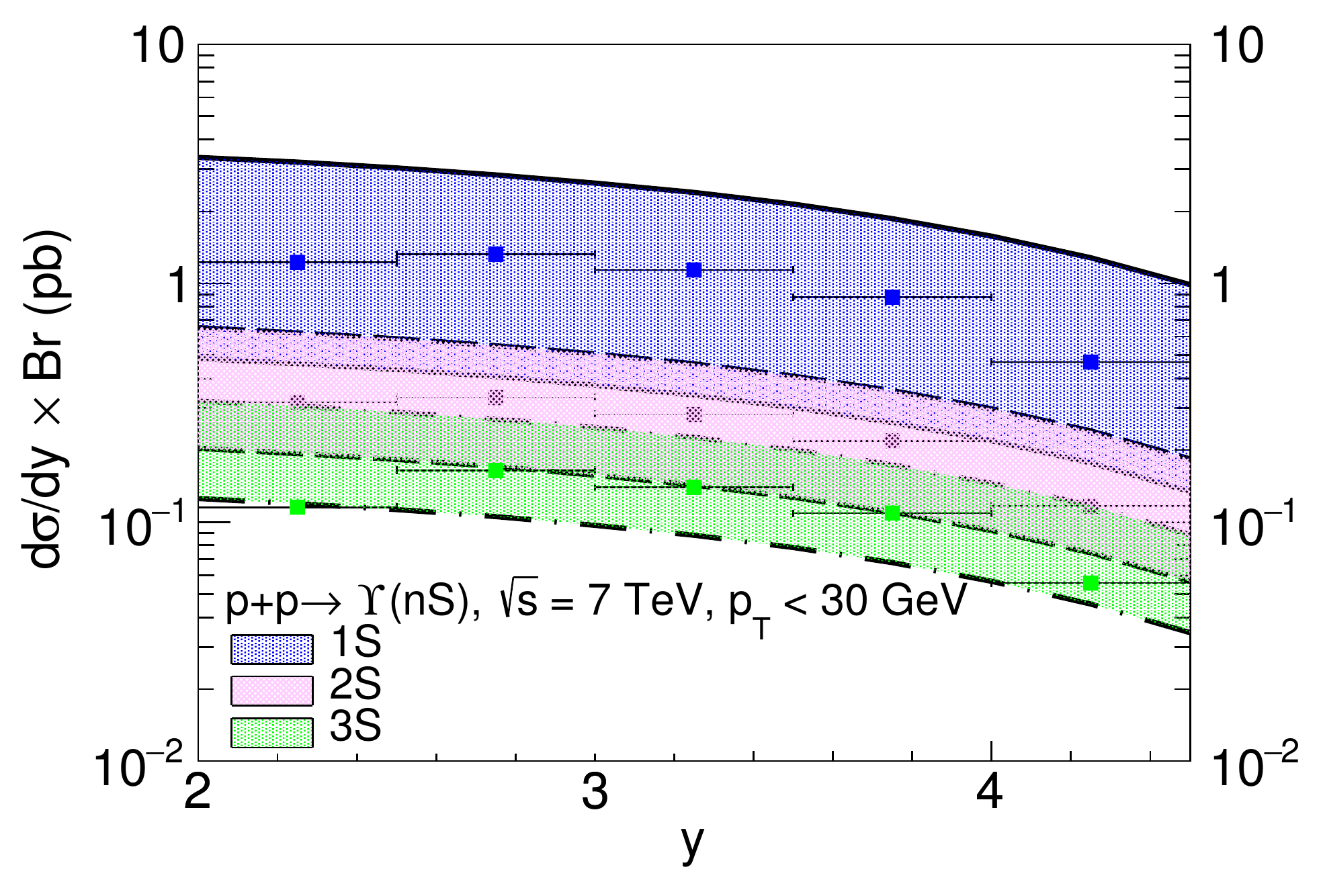}
\caption{(color online) The rapidity dependence of prompt $\varUpsilon$(1S) (blue solid), $\varUpsilon$(2S) (magenta dashed) and $\varUpsilon$(3S) (green dot-dashed) production at $\sqrt{s} = 7$~TeV integrated over $p_T<30$~GeV in the ICEM with combined mass and renormalization scale uncertainties are compared with the LHCb data \cite{Aaij:2015awa}.} \label{LHCb_nS_rapidity}
\end{figure}

We test the universality of $F_{\varUpsilon{\rm(1S)}}$ by comparing the prompt $\varUpsilon$(1S) $p_T$ distribution in the ICEM measured by LHCb \cite{Aaij:2015awa} at $\sqrt{s}=7$~TeV and $2<y<4.5$ in Fig.~\ref{LHCb_1S_pt} and to the prompt $\varUpsilon$(1S) $p_T$ distribution measured by D0 \cite{Abachi:1995tj} at $\sqrt{s}=1.8$~TeV and $|y|<0.5$ in Fig.~\ref{D0_1S_pt}. We again assume the direct production is a constant fraction, $0.71$, of the prompt production to obtain the prompt $\varUpsilon$(1S) cross section. We find the ICEM result agrees with the data for all $p_T$.


\subsubsection{$\varUpsilon$(2S) $p_T$ distribution}

The prompt $\varUpsilon$(2S) $p_T$ distribution at $\sqrt{s} = 7$~TeV is compared to the CMS measurement \cite{Chatrchyan:2013yna} over $|y|<2.4$ in Fig.~\ref{CMS_2S_pt} and the LHCb data \cite{Aaij:2015awa} in $2<y<4.5$ in Fig.~\ref{LHCb_2S_pt}. Here, the direct production cross section is calculated using Eq.~(\ref{pt_rap_cut}) by integrating the pair invariant mass from $M_{\varUpsilon{\rm (2S)}}$ to $2m_{B^0}$ over the rapidity range $|y|<2.4$. Similar to direct $\varUpsilon$(1S), we assume the direct production of $\varUpsilon$(2S) is a constant fraction, 0.73, of the prompt production. We then compare the $p_T$-integrated yield of prompt $\varUpsilon$(2S) with the CMS measurement \cite{Chatrchyan:2013yna}. By matching the $p_T$-integrated yield, we find $F_{\varUpsilon{\rm (2S)}}=0.0144$. We note that $F_{\varUpsilon{\rm (2S)}} \gtrsim F_{\varUpsilon{\rm (1S)}}$, primarily because the integrated mass region is much narrower for $\varUpsilon$(2S) than $\varUpsilon$(1S), a difference of $\sim$0.5~GeV in this case. In the traditional CEM, $F_{\varUpsilon{\rm (2S)}}$ is smaller than $F_{\varUpsilon{\rm (1S)}}$ because the range of integration over the pair invariant mass is the same for all $\varUpsilon$($n$S). We find agreement with the data within the combined uncertainty band constructed by varying the bottom quark mass and the renormalization scale in the ICEM. In both cases, the calculations, with their associated uncertainty bands, are in agreement with the data.

\begin{figure*}
\centering
\begin{minipage}[ht]{0.6873\columnwidth}
\centering
\includegraphics[width=\columnwidth]{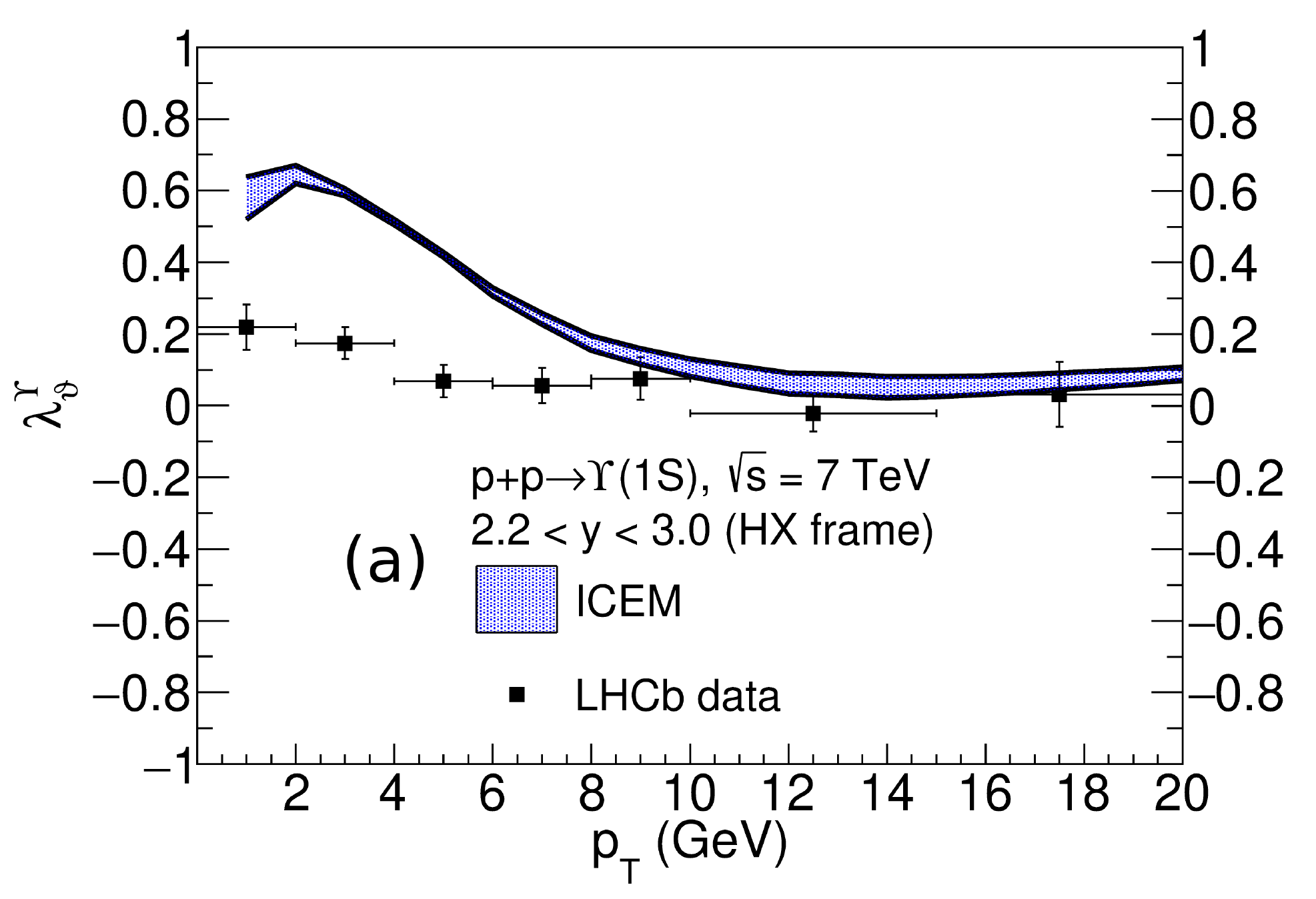}
\end{minipage}%
\begin{minipage}[ht]{0.6873\columnwidth}
\centering
\includegraphics[width=\columnwidth]{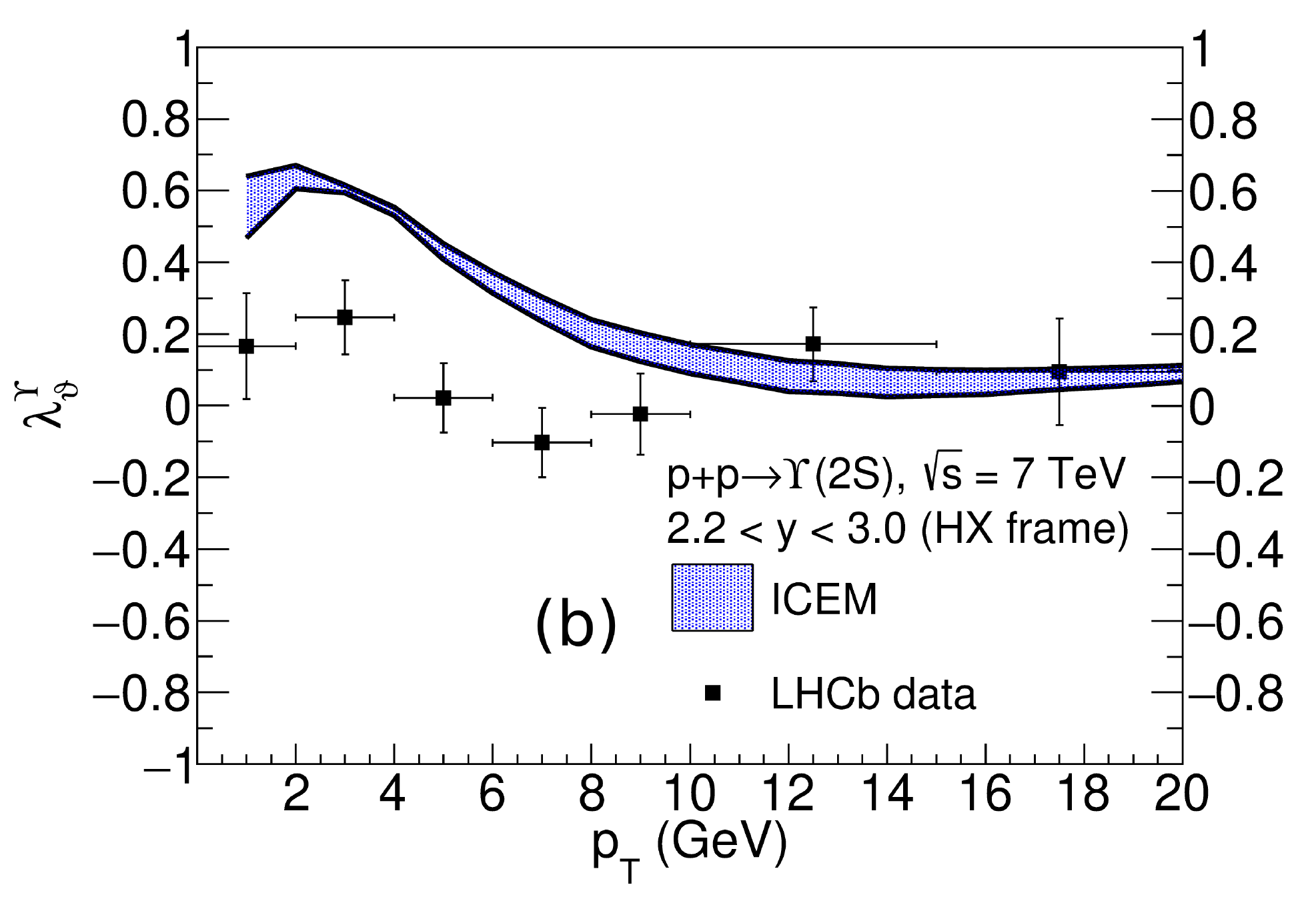}
\end{minipage}
\begin{minipage}[ht]{0.6873\columnwidth}
\centering
\includegraphics[width=\columnwidth]{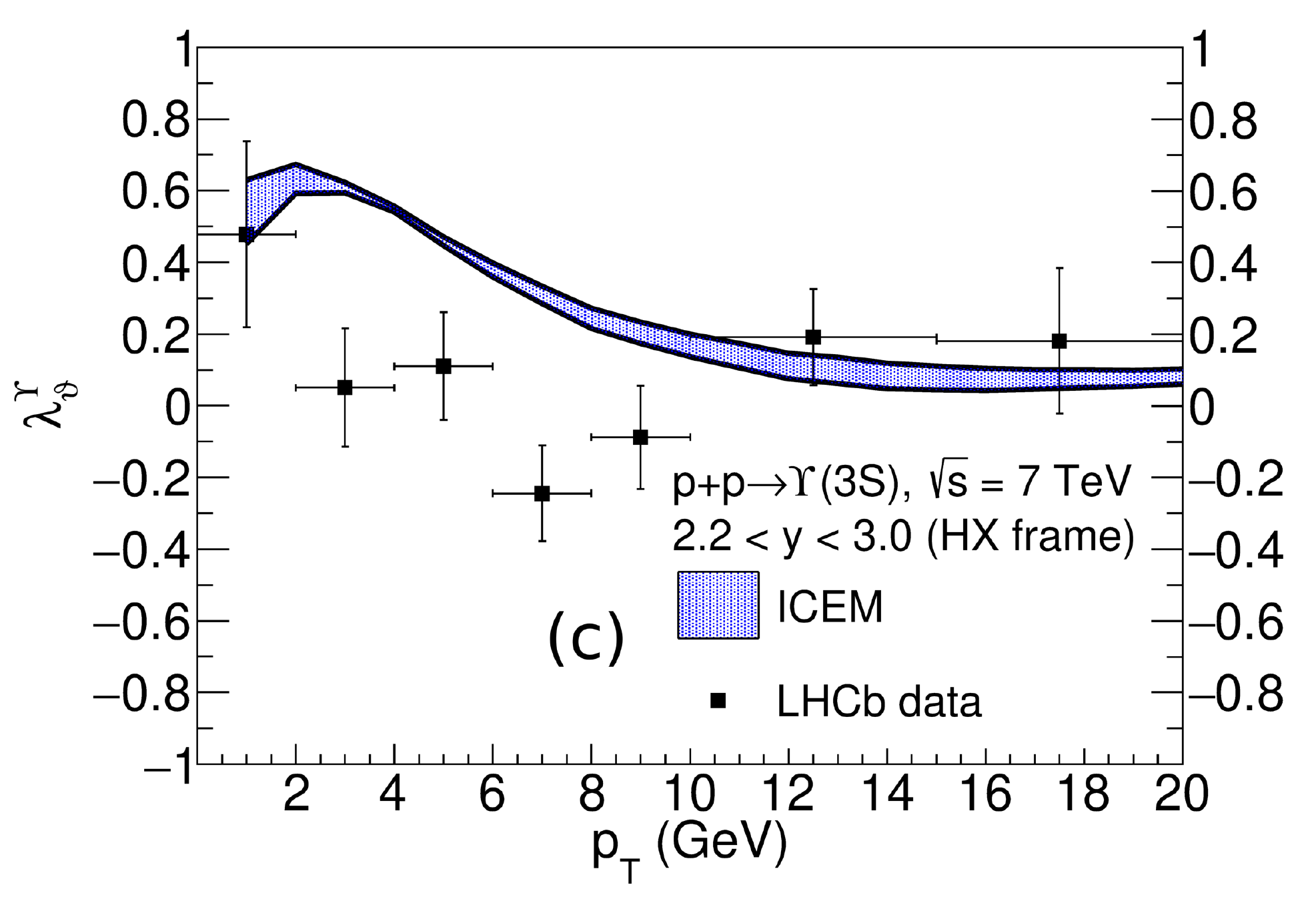}
\end{minipage}
\caption{The $p_T$ dependence of the polarization parameter $\lambda_\vartheta$ for prompt $\varUpsilon$(1S) (a), $\varUpsilon$(2S) (b) and $\varUpsilon$(3S) (c) production in the helicity frame at $\sqrt{s} = 7$~TeV in the ICEM using the ``low $p_T$'' $c_\mathcal{Q}$'s with mass uncertainties are compared to the LHCb data in the range $2.2<y<3$ \cite{Aaij:2017egv}.} \label{LHCb_nS_forward_HX}
\end{figure*}

\begin{figure}[hb]
\centering
\includegraphics[width=\columnwidth]{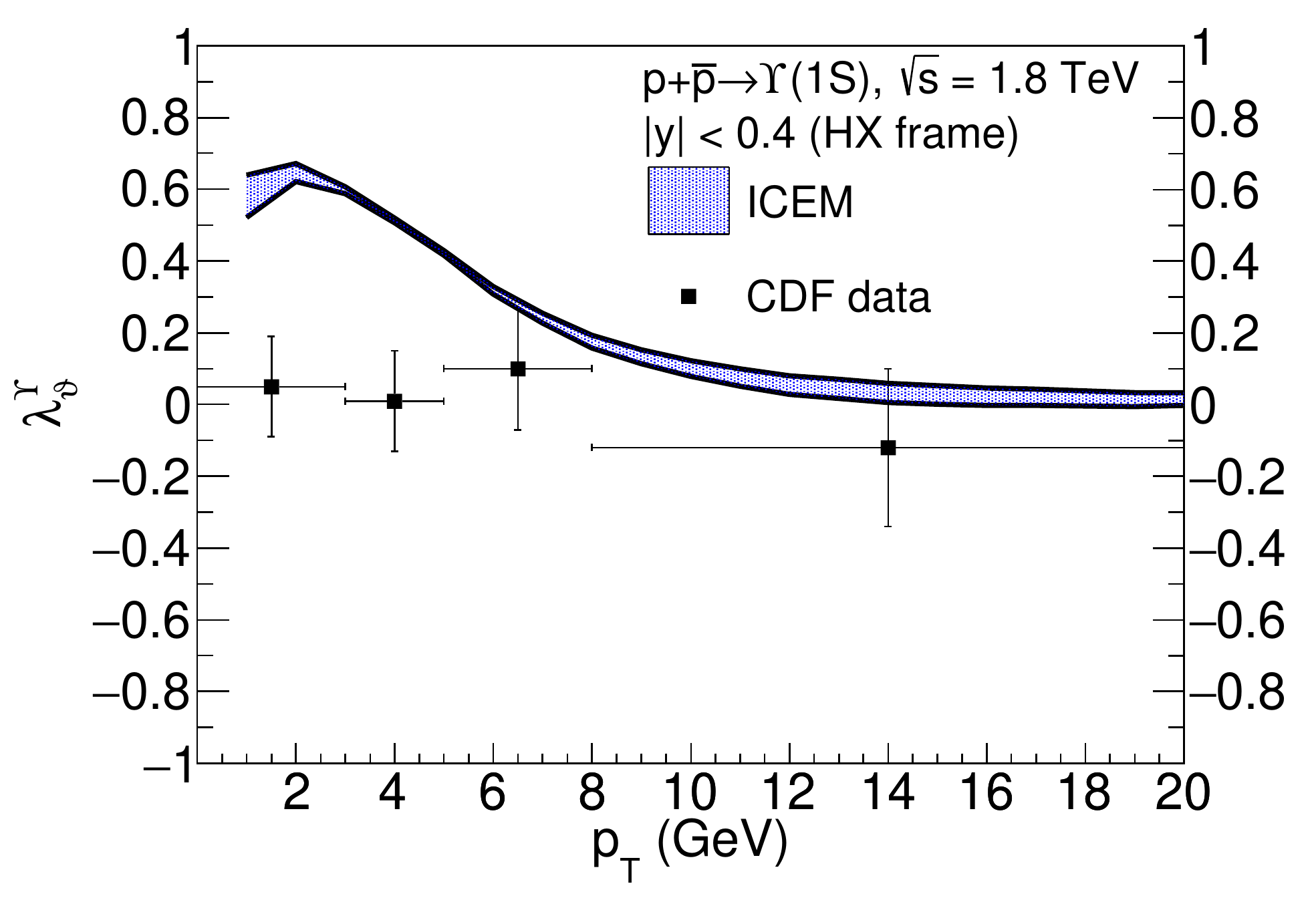}
\caption{The $p_T$ dependence of the polarization parameter $\lambda_\vartheta$ for prompt $\varUpsilon$(1S) production in the helicity frame at $\sqrt{s} = 1.8$~TeV with $|y|<0.4$ in the ICEM using the ``low $p_T$'' $c_\mathcal{Q}$'s \cite{Andronic:2015wma} with mass uncertainties are compared to the CDF data \cite{Acosta:2001gv}.} \label{CDF_1S_HX}
\end{figure}

\subsubsection{$\varUpsilon$(3S) $p_T$ distribution}

The prompt $\varUpsilon$(3S) $p_T$ distribution at $\sqrt{s} = 7$~TeV is compared to the CMS measurements \cite{Chatrchyan:2013yna} over $|y|<2.4$ in Fig.~\ref{CMS_3S_pt} and the LHCb data \cite{Aaij:2015awa} in $2<y<4.5$ in Fig.~\ref{LHCb_3S_pt}. Here, the direct production cross section is calculated using Eq.~(\ref{pt_rap_cut}) by integrating the pair invariant mass from $M_{\varUpsilon{\rm (3S)}}$ to $2m_{B^0}$ over the rapidity range $|y|<2.4$. Similar to direct $\varUpsilon$(1S), we assume the direct production of $\varUpsilon$(3S) is a constant fraction, 0.70, of the prompt production. Therefore, we compare the $p_T$-integrated yield of direct $\varUpsilon$(3S) with the CMS measurement \cite{Chatrchyan:2013yna}. We find $F_{\varUpsilon{\rm (3S)}}=0.00229$. We note that also $F_{\varUpsilon{\rm (3S)}} \gtrsim F_{\varUpsilon{\rm (1S)}}$, because the mass range is still smaller for $\varUpsilon$(3S), a difference of only $\sim0.15$~GeV. Again, in the traditional CEM, $F_{\varUpsilon{\rm (3S)}}$ is smaller than $F_{\varUpsilon{\rm (1S)}}$ and $F_{\varUpsilon{\rm (2S)}}$ because the range of integration over the pair invariant mass is also the same for both $\varUpsilon$(1S) and $\varUpsilon$(3S). There is fair agreement with the data within the combined uncertainty band constructed by varying the bottom quark mass and the renormalization scale in the ICEM. In both cases, the calculations, with their associated uncertainty bands, are in agreement with the data.


\subsubsection{Ratio of $\chi_{\rm b2}$(1P) to $\chi_{\rm b1}$(1P) production}

We now turn to the $p_T$ dependence of the ratio $\chi_{b2}$(1P)/$\chi_{b1}$(1P) as a function of $p_T$. The ratios of direct $\chi_{b2}$(1P) to direct $\chi_{b1}$(1P) at $\sqrt{s}=8$~TeV at central and forward rapidities are presented in Fig.~\ref{chib_ratios}. Direct production is calculated using Eq.~(\ref{pt_rap_cut}) by integrating the pair invariant mass from $M_{\chi_{b1,2}}$(1P) to $2m_{B^0}$ over two rapidity ranges, $|y|<1.5$ and $2<y<4.5$ respectively, in order to compare with existing measurements \cite{Khachatryan:2014ofa,Aaij:2014hla}. As there is not enough information on the feed-down production to $\chi_{b}$, we assume the prompt production of $\chi_{b1,2}$(1P) is approximately the same as the direct production. Since there are no measurements of the absolute $\chi_{b1,2}$(1P) production cross sections, we cannot fix $F_{\chi_{b1,2}{\rm(1P)}}$. Furthermore, the data reports the ratio as a function of the $p_T$ of $\varUpsilon$(1S). To compare our results with the data, we then assume that $p_{T}^{\chi_b}\approx p_{T}^{\varUpsilon{\rm(1S)}}$, not unreasonable since the mass difference between the states is $\sim500$~MeV and the decay photon is soft. Thus the ICEM can only predict the trend of the relative production subject to an overall vertical shift. Similar to the $\chi_{c2}$ to $\chi_{c1}$ ratio in the ICEM \cite{Cheung:2018tvq}, $\chi_{b2}$(1P)/$\chi_{b1}$(1P) becomes constant for $p_T>2 M_{\chi_b}$. However, the relative production decreases with increasing $p_T$ for $p_T < 2M_{\chi_b}$, independent of the rapidity range considered. Our ICEM results only agree with the data in the higher $p_T$ range. This is because the difference between the amplitudes of $\chi_{b1}$ and $\chi_{b2}$ is most apparent at low $p_T$ since the curvature of the distributions changes fastest near the peaks of the distributions. However, the measured relative production is approximately $p_T$ independent at lower $p_T$. We note that the $\chi_{c2}/\chi_{c1}$ ratios presented in Ref.~\cite{Cheung:2018tvq} agreed with the data over the measured $p_T$ range because, in that case, $p_T>>M_{\chi_c}$ over the range of the measurement. However, with the lower $p_T$ range here this condition is not satisfied for $\chi_b$.

\begin{figure*}
\centering
\begin{minipage}[ht]{0.6873\columnwidth}
\centering
\includegraphics[width=\columnwidth]{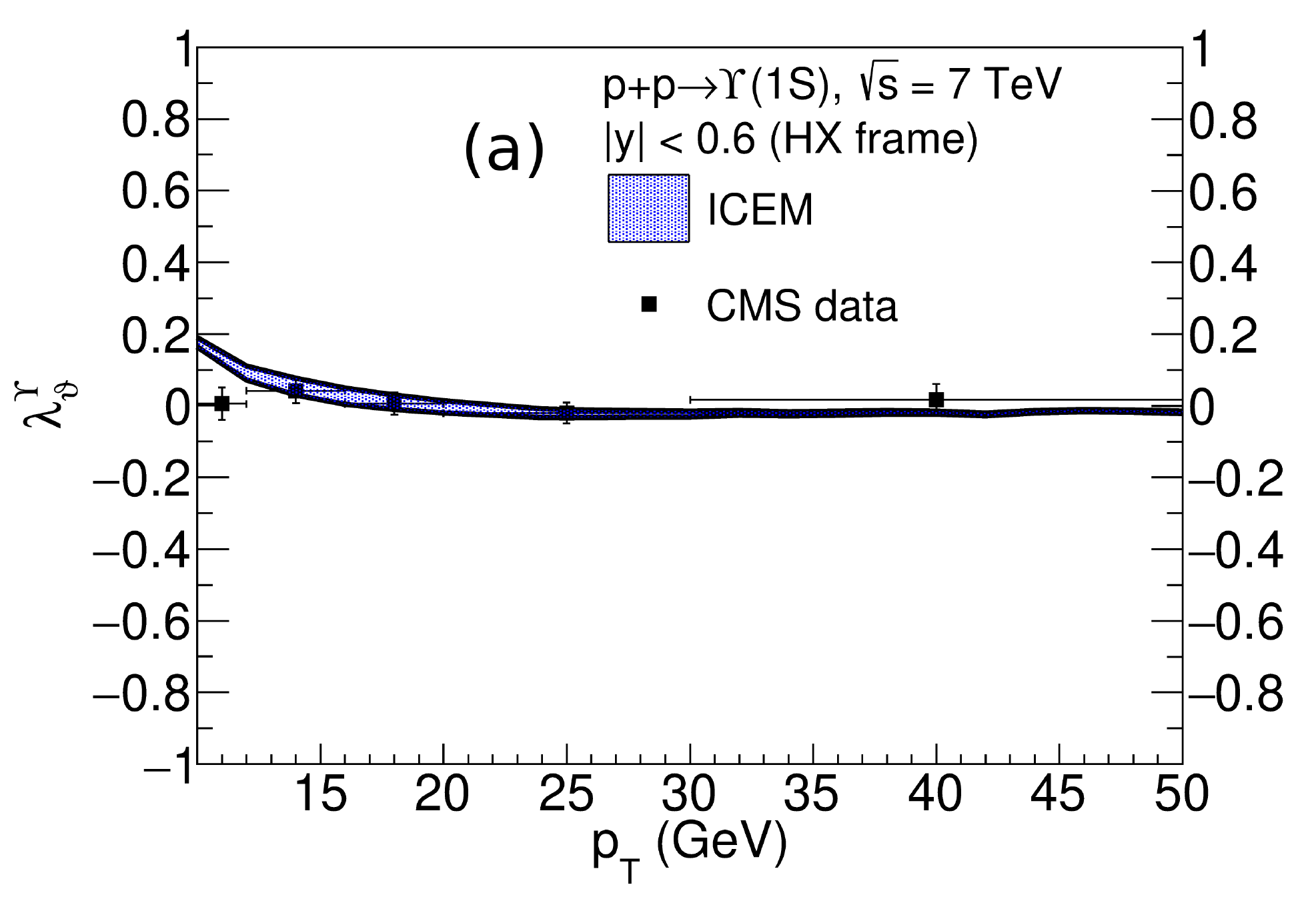}
\end{minipage}%
\begin{minipage}[ht]{0.6873\columnwidth}
\centering
\includegraphics[width=\columnwidth]{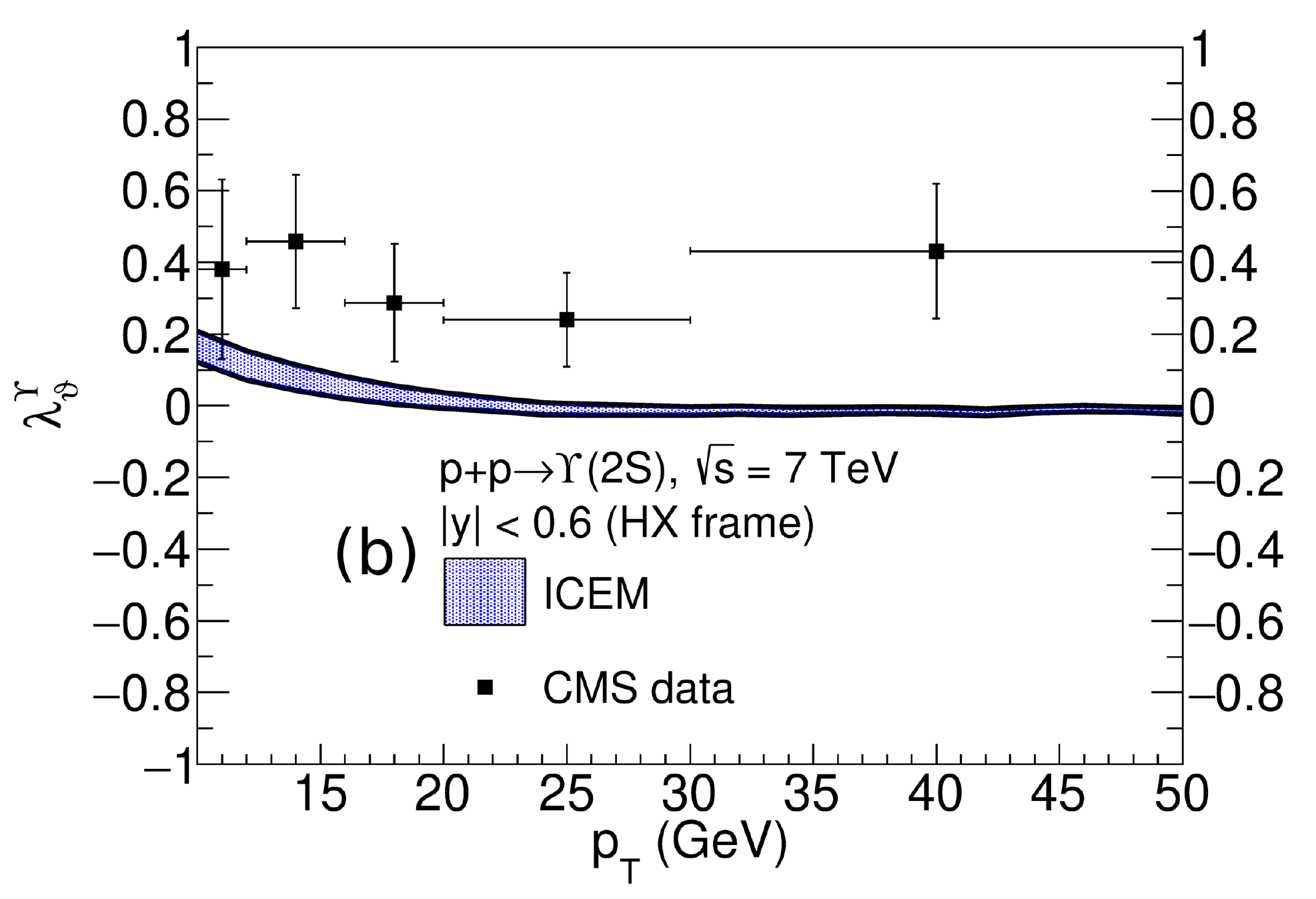}
\end{minipage}
\begin{minipage}[ht]{0.6873\columnwidth}
\centering
\includegraphics[width=\columnwidth]{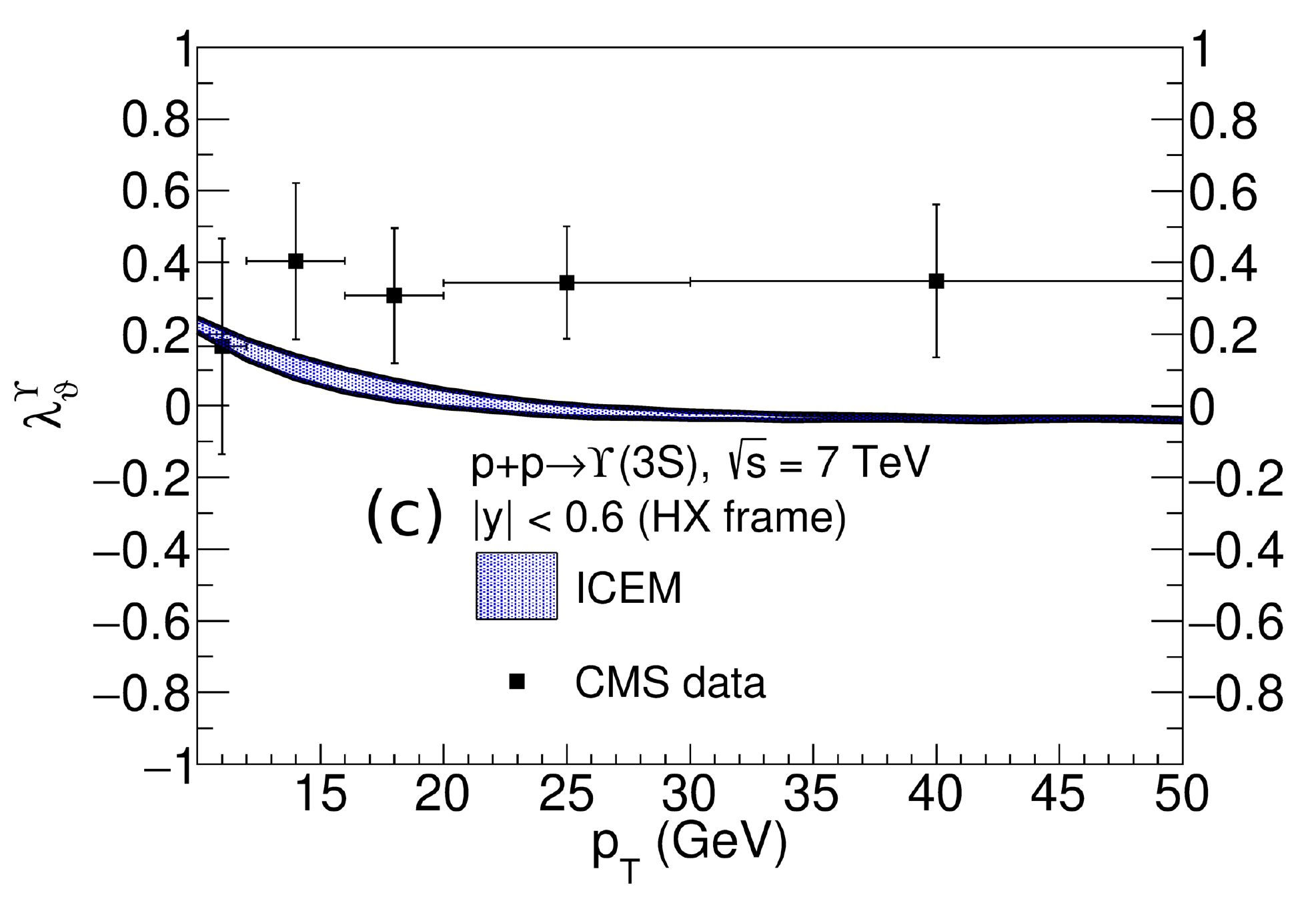}
\end{minipage}
\caption{The $p_T$ dependence of the polarization parameter $\lambda_\vartheta$ for prompt $\varUpsilon$(1S) (a), $\varUpsilon$(2S) (b) and $\varUpsilon$(3S) (c) production in the helicity frame at $\sqrt{s} = 7$~TeV in the ICEM using the ``high $p_T$'' $c_\mathcal{Q}$'s \cite{Andronic:2015wma} with mass uncertainties are compared to the CMS dataat midrapidity in the range $|y|<0.6$ \cite{Chatrchyan:2012woa}.} \label{CMS_nS_central_HX}
\end{figure*}

\subsubsection{$\varUpsilon$($n$S) rapidity distribution}

We now turn to the rapidity dependence of $\varUpsilon$($n$S) production. The rapidity distribution of prompt of $\varUpsilon$($n$S) at $\sqrt{s} = 7$~TeV is shown  in Fig.~\ref{LHCb_nS_rapidity}. The direct production is calculated using Eq.~(\ref{y_pt_cut}) by integrating over the $p_T$ range $0<p_T<30$~GeV. We again assume the direct production of $\varUpsilon$(1S, 2S, 3S) is a constant 71\%, 73\% and 70\% of prompt $\varUpsilon$(1S, 2S, 3S) production respectively. We use the same values of $F_{\varUpsilon{\rm(}n{\rm S)}}$ determined for the $p_T$ distributions to compare the rapidity distribution in the ICEM with the measurement made by the LHCb Collaboration \cite{Aaij:2015awa}. We find the ICEM can describe the LHCb rapidity distribution at $\sqrt{s}=7$~TeV using the $F_{\varUpsilon{\rm(}n{\rm S)}}$ obtained at the same energy by CMS in the central rapidity region.

\subsection{$p_T$ dependence of $\lambda_\vartheta$}

Here, we present the $p_T$ dependence of the polarization parameter $\lambda_\vartheta$ in $p+p$ and $p+\bar{p}$ collisions. Because the polarization parameter is defined as the ratio of polarized to unpolarized cross sections in Eq.~(\ref{mix_upsilon}) and these cross sections depend on $\mu_R$ in the same way, the polarization parameter is independent of the scale choice. Note that $\lambda_\vartheta$ is thus also independent of $\mu_F$. However, the amplitudes themselves are mass dependent so that the polarized to unpolarized ratio in $\lambda_\vartheta$ depends on the bottom quark mass. Thus the only uncertainty on $\lambda_\vartheta$ in our calculation is due to the variation of $m_b$ in the range $4.5 < m_b < 5$~GeV. Therefore, in this section, the uncertainty bands only include the mass variation and the uncertainty in the calculated polarization is reduced relative to those of the yield calculations.

We note that the $J_z$ components of the polarized cross section depend differently on the bottom quark mass. When $p_T \leq M_\mathcal{Q}$, the longitudinally polarized partonic cross section decreases faster with increasing $m_b$ than the transversely polarized partonic cross section in the helicity frame. Thus increasing the bottom quark mass results in more transverse polarization. When $p_T > M_\mathcal{Q}$, the longitudinally-polarized partonic cross section decreases more slowly with increasing $m_b$ than the transversely-polarized partonic cross section. Thus, increasing the bottom quark mass results in more longitudinal polarization. As $p_T \gg \hat{s}$, $\lambda_\vartheta$ becomes insensitive to $m_b$. Thus the uncertainty in $\lambda_\theta$ is narrower at high $p_T$.

Our calculation also depends on the feed-down ratios presented in Table.~\ref{feeddown}, taken from Ref.~\cite{Andronic:2015wma}. Here, ``low $p_T$'' refers to $p_T\lesssim 20$~GeV and ``high $p_T$" refers to $p_T\gtrsim 20$~GeV. We use the ``low $p_T$" ratios to compare our results with LHCb data $(0<p_T<20$~GeV) and the ``high $p_T$" ratios to compare with the CMS data $(10<p_T<50$~GeV).

\subsubsection{prompt $\varUpsilon$($n$S) polarization in $p+p$($\bar{p}$) collisions at low $p_T$}

We present the polarization parameters for prompt $\varUpsilon$(1S) in $p+p$ collisions at $\sqrt{s}=7$~TeV at forward rapidity ($2.2<y<3$) in the helicity frame (HX) in Fig.~\ref{LHCb_nS_forward_HX}. We compare our results with data from the LHCb Collaboration in the forward rapidity region \cite{Aaij:2017egv}. The ICEM polarization of prompt $\varUpsilon$($n$S) in the helicity frame is slightly transverse at low $p_T$ ($p_T<M_{\varUpsilon}$). The result becomes unpolarized for $p_T > M_{\varUpsilon}$. We do not find that the polarization has any significant rapidity dependence. The ICEM polarization agrees with the LHCb data for $p_T > M_{\varUpsilon}$.

We also compare the polarization parameter for prompt $\varUpsilon$(1S) in $p+\bar{p}$ at $\sqrt{s}=1.8$~TeV with the data measured by the D0 Collaboration in the region $|y|<0.4$ \cite{Acosta:2001gv} in the helicity frame, shown in Fig.~\ref{CDF_1S_HX}. We also do not find a strong dependence on $\sqrt{s}$ for the prompt $\varUpsilon$(1S) polarization in the ICEM. The trend in the $p_T$ dependence of the polarization is the same. At the highest $p_T$ bin, the prompt $\varUpsilon$(1S) polarization measured by the D0 Collaboration is slightly longitudinal while still agreeing with the ICEM calculation, which gives an unpolarized result.

We do not find significant differences in the polarizations among the $\varUpsilon$($n$S) states. This is because the calculations of the $\varUpsilon$($n$S) states differ from one another only by the integration limits of the ICEM. Furthermore, the polarization depends only on the ratio of polarized to unpolarized cross sections. Thus there is only a slight difference in polarization whether only direct production is included or if feed down also contributes. Therefore the polarization of $\varUpsilon$($n$S) from $\chi_b$ feed down is similar to that for direct production $\varUpsilon$($n$S) alone. Thus, varying the feed-down ratio, either by adopting the ``high $p_T$'' ratios from Ref.~\cite{Andronic:2015wma} used here or the $p_T$-independent ratios calculated in Ref.~\cite{Digal:2001ue} and used in Ref.\cite{Cheung:2017osx}, changes the polarization by less than 0.05 over all $p_T$. Our results differ from an NLO NRQCD calculation finding that all $\varUpsilon$($n$S) states are unpolarized: $(-0.2< \lambda_\vartheta < 0.2)$ at low $p_T$ \cite{Gong:2013qka}. In their approach, at low $p_T$, the direct $\varUpsilon$($n$S) states are slightly longitudinally polarized while the contribution from $\chi_{b}$ feed down is slightly transverse, resulting in unpolarized prompt production.

\subsubsection{prompt $\varUpsilon$($n$S) polarization in $p+p$($\bar{p}$) collisions at high $p_T$}

We present the polarization parameters for prompt $\varUpsilon$(1S) in $p+p$ collisions at $\sqrt{s}=7$~TeV at central rapidity ($|y|<0.6$) in the helicity frame respectively in Fig.~\ref{CMS_nS_central_HX}. We compare our results with the data from the CMS Collaboration in the central rapidity region \cite{Chatrchyan:2012woa}. The ICEM polarization of prompt $\varUpsilon$ in the helicity frame is near unpolarized at intermediate $p_T$ ($p_T\sim M_{\varUpsilon}$). We see that $\lambda_\vartheta$ becomes unpolarized for $p_T > M_{\varUpsilon}$. The ICEM polarization agrees with the CMS data for $\varUpsilon$(1S) and only agrees with $\varUpsilon$(2S) and $\varUpsilon$(3S) data within 2$\sigma$. We do not find that the polarization has any significant rapidity dependence.

We note that here we have used the ``high $p_T$'' set of feed-down ratios to consider the prompt $\varUpsilon$($n$S) polarization. Although the contribution from direct $\varUpsilon$(1S) to prompt $\varUpsilon$(1S) drops from 71\% to 45\%, the polarization of the prompt production does not change significantly. This is because the polarization of all the bottomonium states below the $B\bar{B}$ threshold are very similar after feed down to prompt $\varUpsilon$($n$S). We note that the polarization at intermediate $p_T$, $p_T\sim15$~GeV, has no significant dependence on the choice of feed-down ratios, as shown in Figs.~\ref{LHCb_nS_forward_HX} and \ref{CMS_nS_central_HX}. The variation of the feed down fractions is negligible compared to the bottom quark mass variation.

\begin{figure*}
\centering
\begin{minipage}[ht]{0.97\columnwidth}
\centering
\includegraphics[width=\columnwidth]{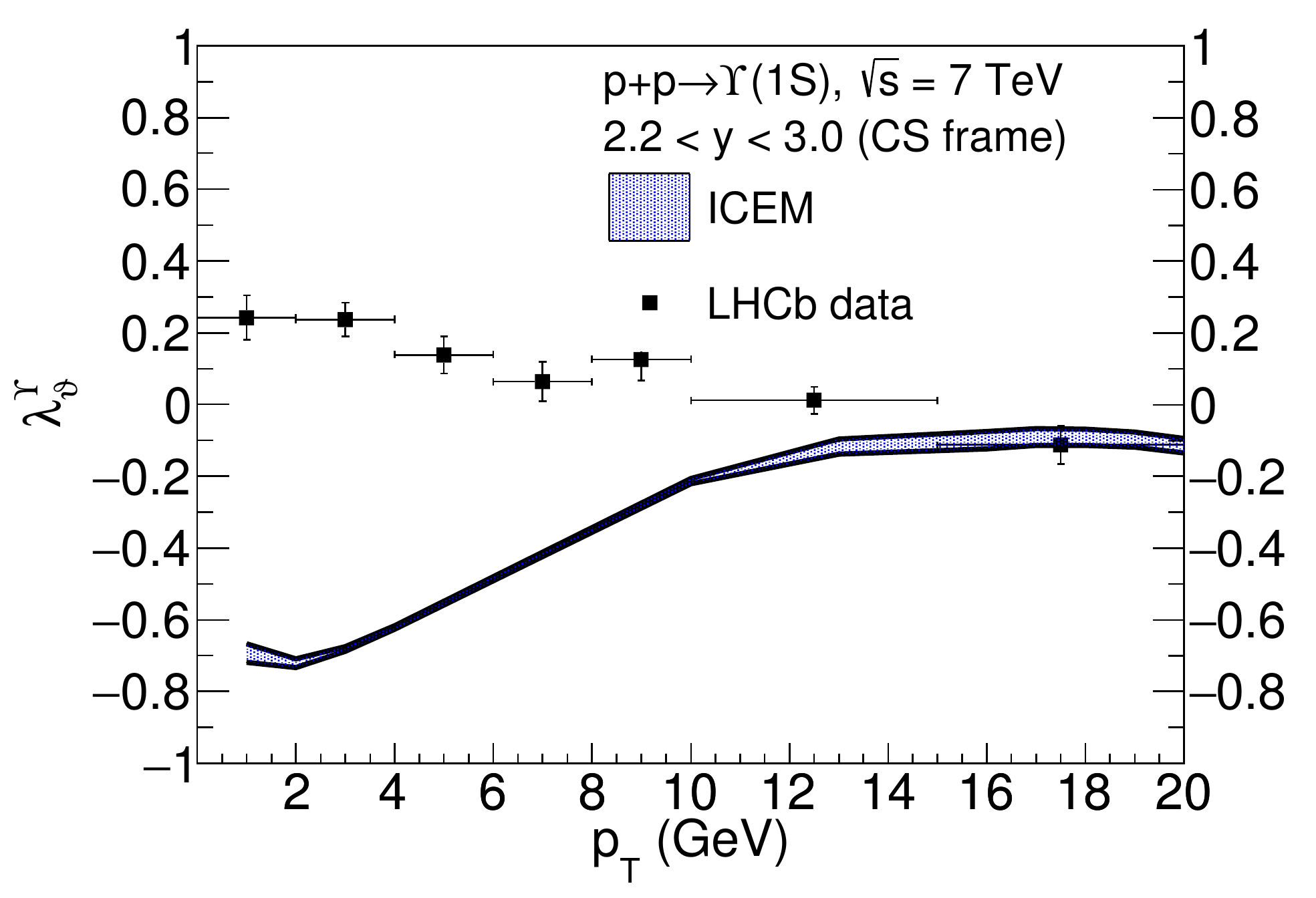}
\caption{The $p_T$ dependence of the polarization parameter $\lambda_\vartheta$ for prompt $\varUpsilon$(1S) production in the Collins-Soper frame at $\sqrt{s} = 7$~TeV and $2.2<y<3$ in the ICEM using the ``low $p_T$'' $c_\mathcal{Q}$'s \cite{Andronic:2015wma} with mass uncertainties are compared to the LHCb data \cite{Aaij:2017egv}.} \label{LHCb_1S_forward_CS}
\end{minipage}%
\hspace{1cm}%
\begin{minipage}[ht]{0.97\columnwidth}
\centering
\includegraphics[width=\columnwidth]{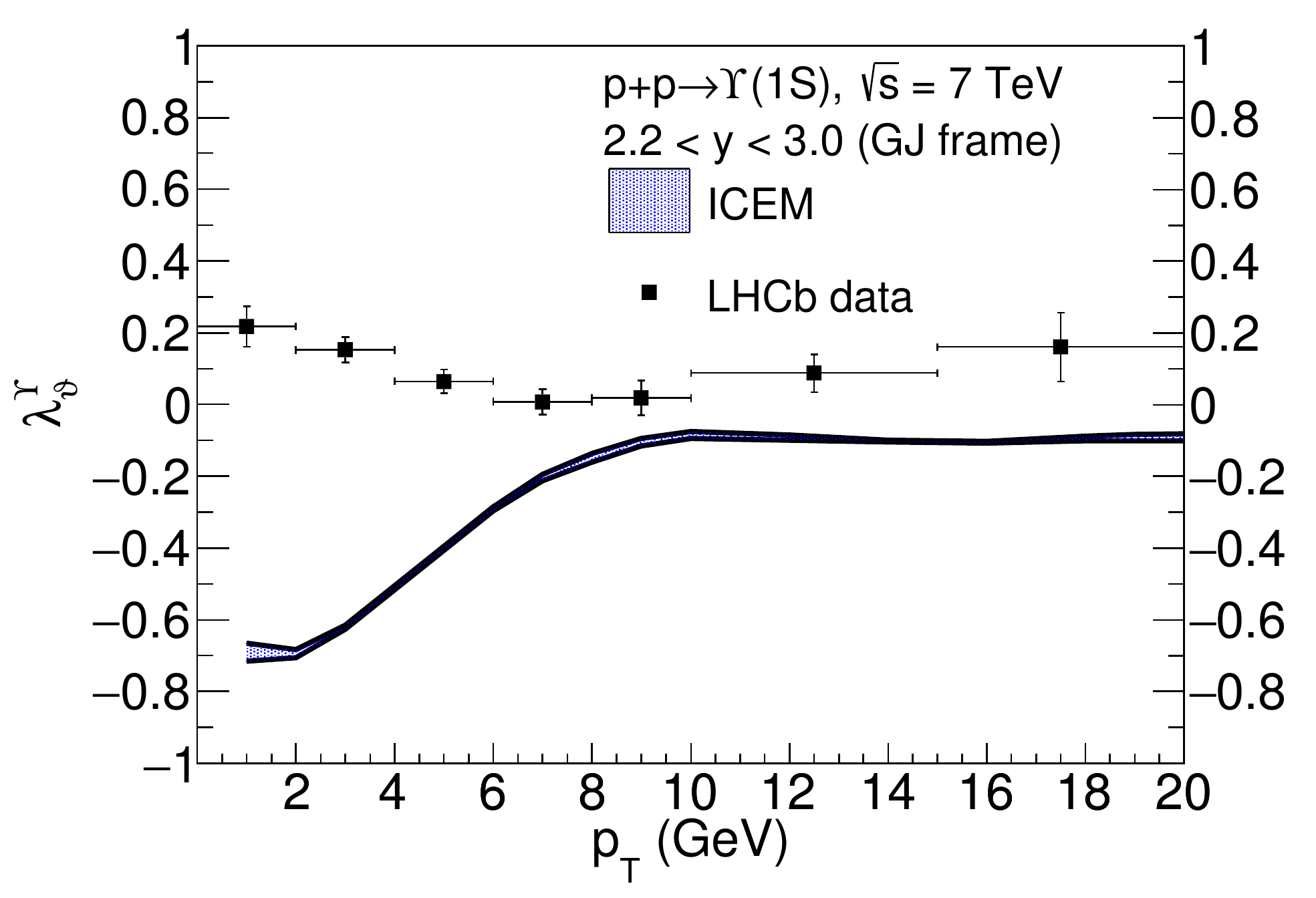}
\caption{The $p_T$ dependence of the polarization parameter $\lambda_\vartheta$ for prompt $\varUpsilon$(1S) production in the Gottfried-Jackson frame at $\sqrt{s} = 7$~TeV and $2.2<y<3$ in the ICEM using the ``low $p_T$'' $c_\mathcal{Q}$'s \cite{Andronic:2015wma} with mass uncertainties are compared to the LHCb data \cite{Aaij:2017egv}.} \label{LHCb_1S_forward_GJ}
\end{minipage}
\end{figure*}

Similar to our results at low $p_T$, we do not find significant differences in polarizations among the $\varUpsilon$($n$S) states. Our results differ from an NLO NRQCD calculation finding that the polarization at $p_T \gtrsim 20$~GeV is more transverse for higher mass bound states, saturating at $\lambda_\vartheta \sim 0.2$, $\sim 0.4$ and $\sim 0.9$ for $\varUpsilon$(1S), $\varUpsilon$(2S) and $\varUpsilon$(3S) respectively \cite{Gong:2013qka}. The significant transverse polarization of $\varUpsilon$(3S) in their approach is due to the fact that the polarization is calculated without the contribution from $\chi_{b}$ feed-down production.

\subsection{Frame dependence of $\lambda_\vartheta$}

We now turn to the frame dependence of our 7~TeV results. We calculate the polarization parameter in $p+p$ collisions at $\sqrt{s}=7$~TeV in the same kinematic region as presented in Fig.~\ref{LHCb_nS_forward_HX} in both the Collins-Soper and the Gottfried-Jackson frames, shown in Figs.~\ref{LHCb_1S_forward_CS} and \ref{LHCb_1S_forward_GJ} respectively. Since the polarization axes in the helicity frame and the Collins-Soper frame are always perpendicular to each other in $\mathcal{O}(\alpha_s^2)$ kinematics, the polarization in the Collins-Soper frame is opposite to that in the helicity frame in the ICEM. Therefore, at low $p_T$, where the $\varUpsilon$(1S) is predicted to be slightly transverse in the helicity frame, it is predicted to be slightly longitudinal in the Collins-Soper frame. For $p_T>M_{\varUpsilon}$, $\lambda_\vartheta$ is predicted to be unpolarized in both frames. We only find agreement with the data in the Collins-Soper frame for the highest $p_T$ bin. When $p_T \ll m_T$, the angle between the polarization axes in the Gottfried-Jackson frame and that in the Collins-Soper frame is small. As $p_T$ increases, the polarization axis in the Gottfried-Jackson frame becomes collinear with that in the helicity frame. Therefore, the polarization calculated in the Gottfried-Jackson frame is opposite to that in the helicity frame at low $p_T$ and thus similar to that in the Collins-Soper frame. However, as $p_T$ increases, the polarization in the Gottfried-Jackson frame should asymptotically approach the polarization in the helicity frame. Since $\lambda_\vartheta$ is unpolarized in the helicity frame in the high $p_T$ limit, the ICEM polarization becomes frame independent in this limit. We find the ICEM polarization agrees with the data in all frames at high $p_T$ but does not agree with the low $p_T$ data where the frame dependence is most significant.

\section{Conclusions}

We have presented the transverse momentum distributions of the prompt $\varUpsilon$($n$S) cross section as well as the the polarization of prompt $\varUpsilon$($n$S) production in $p+p$ and $p+\bar{p}$ collisions in the improved color evaporation model in the $k_T$-factorization approach. We compared the $p_T$ dependence to data at collider energies. We also presented the ratio $\chi_{b2}$(1P)/$\chi_{b1}$(1P) as a function of $p_T$ at $\sqrt{s}=8$~TeV. We find prompt $\varUpsilon$($n$S) production to be unpolarized at $p_T\gtrsim M_{\varUpsilon}$, independent of frame. We do not observe any rapidity or energy dependence in the polarization in the ranges considered.

Since our calculation of the matrix elements is leading order in $\alpha_s$, we expect improvements when we calculate the cross section to $\mathcal{O}(\alpha_s^3)$ in a future publication. 


\section{Acknowledgments}
We thank B.~Kniehl for the initiation of and encouragement throughout this project. This work was performed under the auspices of the U.S. Department of Energy by Lawrence Livermore National Laboratory under Contract No. DE-AC52-07NA27344 and supported by the U.S. Department of Energy, Office of Science, Office of Nuclear Physics (Nuclear Theory) under Contract No. DE-SC-0004014.


\end{document}